\documentclass[aps,prb,twocolumn,showpacs,amsmath]{revtex4}
\usepackage{bm}
\usepackage{graphicx}
\newcommand{\mkright}{draft (\today)} 

\renewcommand{\Im}{\mbox{Im}\,}

\begin{document}

\title{Predicted field-dependent increase of critical currents in asymmetric superconducting nanocircuits}

\author{John R. Clem}
\affiliation{%
	Ames Laboratory--DOE and Department of Physics and Astronomy, 
	Iowa State University, Ames Iowa 50011, USA}

\author{Yasunori Mawatari}
\affiliation{%
	National Institute of Advanced Industrial Science and Technology (AIST), 
	Tsukuba, Ibaraki 305--8568, Japan}

\author{G. R. Berdiyorov}
\affiliation{Departement Fysica, Universiteit Antwerpen, Groenenborgerlaan 171, B-2020 Antwerpen, Belgium}

\author{F. M. Peeters}
\affiliation{Departement Fysica, Universiteit Antwerpen, Groenenborgerlaan 171, B-2020 Antwerpen, Belgium}

\date{\today}

\begin{abstract} 

The critical current of a thin superconducting strip of width $W$ much larger than the Ginzburg-Landau coherence length $\xi$ but much smaller than the Pearl length $\Lambda = 2 \lambda^2/d$ is maximized when the strip is straight with defect-free edges. When a perpendicular magnetic field is applied to a long straight strip, the critical current initially decreases linearly with $H$ but then decreases more slowly with $H$ when vortices or antivortices are forced into the strip.   However, in a superconducting strip containing sharp 90-degree or 180-degree turns,  the zero-field critical current at $H=0$ is reduced because vortices or antivortices are preferentially nucleated at the inner corners of the turns, where current crowding occurs.  Using both analytic London-model calculations and time-dependent Ginzburg-Landau simulations, we predict that in such asymmetric strips the resulting critical current can be {\it increased} by applying a perpendicular magnetic field that induces a current-density contribution opposing  the applied current density at the inner corners.  This effect should  apply to all turns that bend in the same direction.  
\end{abstract}


\maketitle

\section{Introduction
\label{Sec_Intro}}

The critical current of superconductors is of great interest for both fundamental and practical reasons.  On the fundamental side, it is of interest to know the precise mechanism by which the superconducting state breaks down in  high applied electrical currents.   On the practical side, numerous applications of superconductivity require the operating current to be as high as possible without exceeding the critical current.  In this paper we focus on  thin and narrow strips, where it  is known that the critical current is dominated by the edge-pinning critical current,\cite{Tahara90,Jones10,Elistratov02} which is generally  much larger than the bulk-pinning critical current and can even approach the Ginzburg-Landau depairing critical current.\cite{Tinkham96} Pinning in thin films by the edge barrier is closely related to pinning in bulk superconductors  by the Bean-Livingston barrier,\cite{Bean64,Clem74} a barrier against vortex entry that permits high currents to be carried by the surface in parallel applied magnetic fields. Because in bulk superconductors this vortex-entry barrier depends upon both the applied field angle and the surface quality, and because there is practically no exit barrier, some experiments  have found that the critical current depends upon the current direction, leading to partial rectification\cite{Swartz67} or diode effect.\cite{Harrington09,Carapella12}  Theoretical studies have examined how the quality of the surface can alter the Bean-Livingston barrier in bulk superconductors in parallel fields.\cite{Vodolazov00,Aladyshkin01,Hernandez02,Vodolazov03}    Several experimental and theoretical studies of thin films\cite{Grigorieva04,Vodolazov05a,Berdiyorov05,Kuroda10} have examined the role of edge pinning and found vortex nucleation preferentially occurring at defects along the edge or at sharp corners where current crowding\cite{Friesen01,Brandt05,Doenitz06} occurs.  An increase or decrease in the critical current (a diode effect) has been found experimentally to depend upon the sign of a magnetic field parallel to the surface of a superconducting strip with a magnetized magnetic strip on top, an effect the authors explained chiefly in terms of an edge-barrier effect.\cite{Vodolazov05b}
Roughening one of the edges of a straight strip, thereby producing an asymmetric barrier, has been predicted\cite{Vodolazov05c} to lead to rectification upon application of a perpendicular magnetic field.

A recent study\cite{Clem11} has presented calculations of the critical currents in thin superconducting strips with sharp right-angle turns, 180-degree turnarounds, and more complicated geometries, where all the line widths $W$ were much less than the Pearl length\cite{Pearl64} $\Lambda = 2 \lambda^2/d$ (film thickness $d$, London penetration depth $\lambda$, $d < \lambda$) but much greater than the Ginzburg-Landau coherence length $\xi$.  That study, in which the critical current was defined as the current at which the Gibbs free-energy barrier against vortex nucleation is reduced to zero,  showed that current crowding, occurring whenever the current rounds a turn, reduces the critical current below the value it would have in a straight strip with defect-free edges.

In this paper we extend the work of Ref.\ \onlinecite{Clem11} to consider the effect of an applied perpendicular magnetic field $H$ upon the critical current.  In thin and narrow straight superconducting strips for which $\xi \ll W \ll \Lambda$ and the edge barrier to the nucleation of vortices or antivortices dominates the critical current,  the critical current $I_c(H)$ initially decreases linearly with $H$.\cite{Bulaevskii11a}  When $H= 0$, the critical current is determined by the condition that the current density at the strip's edge reduces  to zero the Gibbs free-energy  barrier against nucleation of a vortex on one edge or an antivortex at the opposite edge.  When $H >0$, the current induced by the applied field is in the same direction as the applied current density at one edge but in the opposite direction at the other edge.  Thus the effect of the applied field is either to increase the critical current for the nucleation of a vortex  but to decrease the critical current for the nucleation of an antivortex or to increase the critical current for the nucleation of an antivortex  but to decrease the critical current for the nucleation of a vortex.  In either case, $I_c(H)$ is decreased as $H$ increases, because either vortices or antivortices are nucleated at a lower value of the applied current.  The linear decrease with $H$ changes to a slower rate of decrease at higher fields, when vortices or antivortices are forced to remain in the strip.

If the superconducting strip is {\it asymmetric}, such as making a bend to  the left, when $H= 0$, the critical current is determined solely by the condition that the current density around the bend reduces to zero the Gibbs free-energy barrier against nucleation of a vortex at the inner corner of the bend.  When $H >0$, if the current induced by the applied field is in the same direction as the applied current density at the inner corner, vortex nucleation occurs at a lower value of the applied current and $I_c(H)$ {\it decreases} with $H$.  However, if the  current induced by the applied field opposes the applied current density, vortex nucleation occurs at a higher value of the applied current and $I_c(H)$ {\it increases} with $H$.

In Sec. \ref{StraightStrip}, we discuss the behavior of the critical sheet current $K_c(H_z) = I_c(H_z)/W$  vs applied field $H_z$ in a long  thin and narrow straight superconducting strip of width $W$.  In Sec.\ \ref{Turns}, we consider the behavior of  long thin and narrow strips of width $W$ but with sharp turns       in the middle.  We show that the critical current at zero applied field is reduced below the corresponding critical current of a straight strip with no turn, but we predict that its critical current can be {\it increased} by applying a magnetic field of the right polarity.  In Sec. \ref{TDGL}, we report simulations using time-dependent Ginzburg-Landau equations that generally confirm these predictions.  In Sec.\ \ref{Sec_conclusion}, we summarize our results and discuss possible applications in superconducting nanowire single-photon detectors.  The Appendix contains details of how to calculate the functions needed in Sec.\ \ref{Turns}.

\section{$I_c(H_z)$ for a straight strip
\label{StraightStrip}} 

Consider a superconducting strip of London penetration depth $\lambda$,  thickness $d$ ($d < \lambda$), two-dimensional screening length (Pearl length\cite{Pearl64}) $\Lambda = 2 \lambda^2/d$, width $W$ ($W \ll \Lambda$), and Ginzburg-Landau coherence length ($\xi \ll W$),  centered on the $xy$ plane in the region $0 < y < W$. A current $I$ flows in the $x$ direction. Because  $W \ll \Lambda$, the corresponding sheet-current density $\bm K_I = \hat x K_I = \hat x I/W$ is very nearly independent of $y$.\cite{Clem11}  In the presence of an applied magnetic field $\bm H = \hat z H_z$, the magnetic field penetrates very nearly uniformly through the strip.  The corresponding $H$-induced sheet-current density $\bm K_H = \hat x K_{Hx}(y)$ can be calculated from the London equation $\bm 
K = -(2/\mu_0 \Lambda)[\bm A +(\phi_0/2\pi)\nabla \gamma],$ where $\bm A$ is the vector potential ($\bm B = \nabla \times \bm A$), $\phi_0 = h/2e$ is the superconducting flux quantum, and $\gamma$ is the phase of the order parameter:  $K_{Hx}(y) = (2 H_z/\Lambda)(y - W/2)$.

We next  use a London-model description of vortices to estimate the critical current when the critical current due to the edge barrier greatly exceeds the critical current due to bulk pinning.  We start by writing down the Gibbs free energy $G$ of a vortex nucleating at $y = W$ or an antivortex nucleating at $y = 0$, denoting $\delta$  as the distance from the  edge ($\delta =  W- y$ for the vortex, which carries magnetic flux in the $z$ direction, and $\delta = y$ for the antivortex, which carries magnetic flux in the $-z$ direction):\cite{Clem11,Kogan94,Maksimova98,Kuit08} 
\begin{eqnarray}
G&=&\frac{\phi_0^2}{2\pi \mu_0 \Lambda}\ln\Big[\frac{2W}{\pi \xi}\sin\Big(\frac{\pi \delta}{W}\Big)\Big]-\phi_0 K_I \delta \nonumber \\
&&\mp\frac{\phi_0H_z}{\Lambda}\delta(W-\delta),
\label{Ggeneral}
\end{eqnarray}
where the first term is the self-energy of the vortex or antivortex, accounting for its interactions with an infinite set of images, the second term is the negative of the work done by the source of the applied current as the vortex or antivortex moves from the edge to the coordinate $\delta$, and the third term is the negative of the work done by the source of the applied magnetic field as  the vortex (upper sign) or antivortex (lower sign) moves  from the edge to the coordinate $\delta$.  Note that $G$ is minimized for the vortex when $H_z$ is positive and for the antivortex when $H_z$ is negative.

As in Ref.\ \onlinecite{Clem11}, to estimate the Gibbs free energy barrier in different geometries, in Eq.\ (\ref{Ggeneral}) and later in this paper we use the simplifying assumptions of a London-model vortex and its neglect of the vortex-core energy, which are known to lose accuracy when the vortex is close to the sample edge. Although the numerical factors in our expressions for the critical current therefore are probably not accurate, we expect the qualitative behavior of their geometry dependence to be correct.

\subsection{Linear behavior for small $H$\label{linear}}

To calculate the nucleation of a vortex or antivortex when $\xi \ll W$, we need to examine the behavior of $G$ only for small values of $\delta$, for which, to good approximation,
\begin{equation}
G = \frac{\phi_0^2}{2\pi \mu_0 \Lambda}\ln\Big(\frac{2\delta}{\xi}\Big)-\phi_0 K_I \delta \mp \phi_0 (H_z W/\Lambda)\delta,
\label{Gsmalldelta}
\end{equation}
where the upper (lower) sign holds for the nucleation of a vortex (antivortex).
The free-energy barrier occurs at $\delta = \delta_b$.  Setting $\partial G/\partial \delta = 0$ there, we obtain
\begin{equation}
\delta_b = \frac{\phi_0}{2\pi \mu_0 \Lambda (K_I \pm H_zW/\Lambda)},
\label{deltab}
\end{equation}
which describes a force balance between the repulsive Lorentz force $\phi_0 (K_I \pm H_zW/\Lambda)$ and the attractive force of the nearest image $\phi_0^2/2\pi\mu_0 \Lambda \delta$. 
Setting $G=0$ at $\delta_b$ yields  $\delta_b = \delta_c= e\xi/2 = 1.36\xi \ll W$ and the critical sheet current $K_I = K_c(H_z)$, where
\begin{eqnarray}
K_c(H_z)& =& K_{cs} \mp H_z W/\Lambda, \label{KcstripH}\\
K_{cs}& =&\frac{\phi_0}{e \pi\mu_0\xi\Lambda} ,
\label{Kcstrip0}
\end{eqnarray}
$e=2.718...$ is Euler's number and $K_{cs}$ denotes the critical sheet current for a long straight strip in the absence of an applied field. 
Note that for the nucleation of a vortex (antivortex) the critical sheet current is decreased (increased) for positive $H_z$, because at the nucleation point the $H$-induced current density is in the same (opposite) direction as the applied current density.

\begin{figure}
\includegraphics[width=8cm]{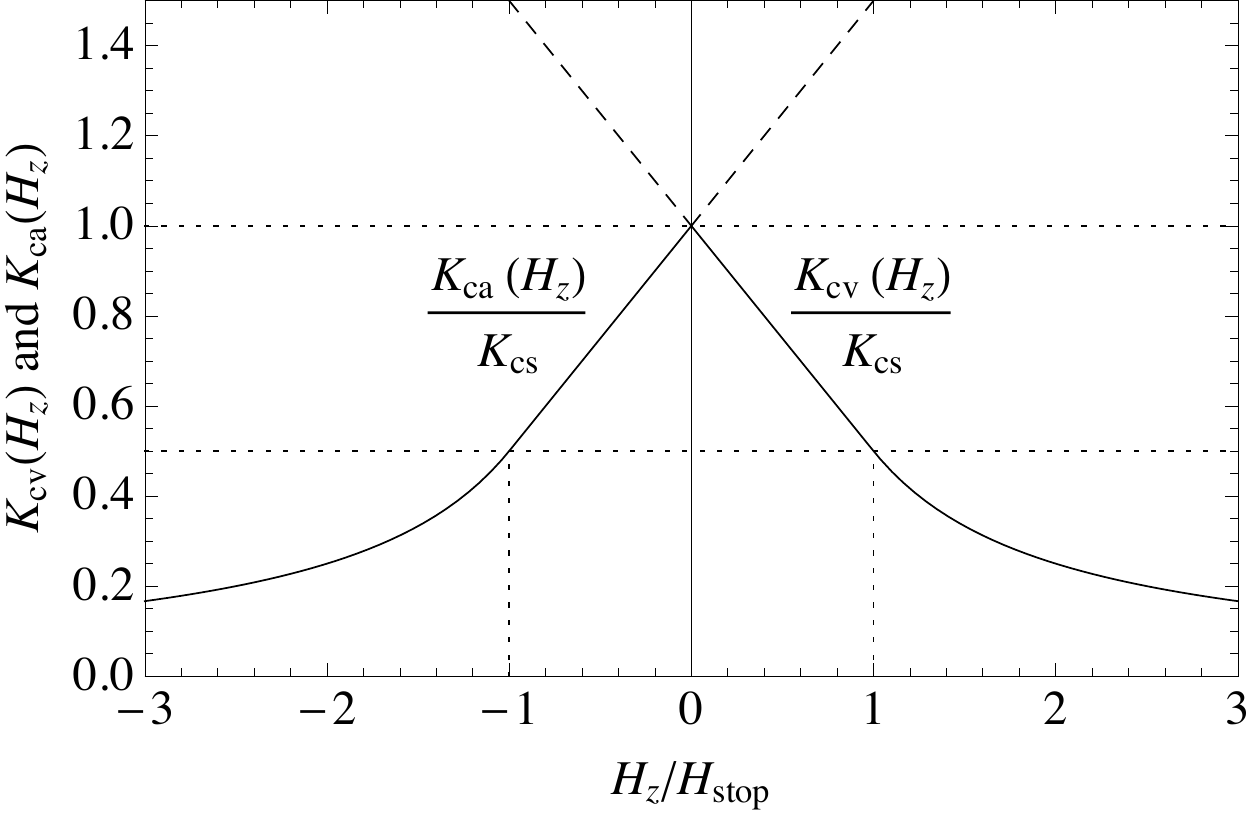}
\caption{%
Critical sheet current for vortices $K_{cv}(H_z)$ and antivortices $K_{ca}(H_a)$ (normalized to $K_{cs}$) vs $H_z$ (normalized to $H_{\rm stop}$).}
\label{Kcv&KcaFig}
\end{figure}

As the applied sheet current $K_I$ is increased, a voltage appears along the length when the smaller of the two critical sheet currents in Eq.\ (\ref{KcstripH}) is reached.  For $H_z > 0$ ($H_z < 0$) the critical current is reached when conditions are favorable for vortices (antivortices) to be nucleated.  Because of this symmetry, with identical barriers for vortices or antivortices on opposite sides of the strip, the critical sheet current for small $H =|H_z|$ is therefore  $K_c(H) = K_{cs} -H W/\Lambda$.  This result also can be written as $K_c(H) = K_{cs}[1- (eW/2\xi)(H/H_{c2})]$, where $H_{c2}=\phi_0/2\pi \mu_0 \xi^2$.  See Fig.\ \ref{Kcv&KcaFig}.

\subsection{Behavior for large $H$\label{nonlinear}}

The linear decrease of $K_{cv}(H_z)=K_{cs} - H_z W/\Lambda$ for the nucleation of vortices (hence the subscript v) given in Eq.\ (\ref{KcstripH}) applies only for relatively small values of $H_z$.  At the critical current the nucleating vortex must be able to travel all the way across the strip and annihilate with its image on the opposite side.  However, it can be seen from Eq.\ (\ref{Ggeneral}) that large values of $H_z$ can produce a free-energy minimum for $\delta < W$.  Accordingly, as $H_z$ increases, there is a special value of $H_z$, which we denote as $H_{\rm stop}$, at which 
this free-energy minimum first appears.  A nucleating vortex therefore comes to a stop at the distance $\delta_{\rm min}$ where the minimum occurs.  When $\xi/W \ll 1$, $\delta_{\rm min}=W(1-\sqrt{e\xi/2W})$, which is very close to the edge where the vortex would have annihilated, and the value of $H_{\rm stop}$ is given to good approximation by 
\begin{equation} 
H_{\rm stop} = \frac{\Lambda}{2W}K_{cs} =\frac{\phi_0}{2\pi e \mu_0 W \xi}=\frac{\xi}{eW}H_{c2}.
\end{equation}  
The linear decrease of $K_{cv}(H_z)$ therefore ceases at $H_z = H_{\rm stop}$, where (to good approximation when $\xi \ll W$)
\begin{equation}
K_{cv}(H_{\rm stop}) =K_{cs}/2.  
\end{equation}

For $H_z > H_{\rm stop}$, $K_{cv}$ decreases more slowly with $H_z$, because the vortices that have stopped within the strip generate an additional current density opposing the applied current density where vortices are nucleated.  To calculate $K_{cv}(H_z)$ accurately becomes a complicated numerical problem, because one must take into account the effect of all the vortices that temporarily reside in their local Gibbs free-energy minima.  

A crude estimate of the field dependence of $K_{cv}(H_z)$ for $H_z > H_{\rm stop}$ can be obtained using an approach  analogous to that used in Ref.\ \onlinecite{Benkraouda98}, in which the field-dependent critical current due to surface barriers was calculated for superconducting strips of width $W \gg \Lambda$, the limit opposite to that of interest to us here. We consider the behavior just below the critical current in a field  $H_z > H_{\rm stop}$ when  an array of vortices is held in place by the applied field.  Our approximation is to replace the discrete vortex array by a stationary distribution of vortices of uniform density $n_v=\mu_0 H_z/\phi_0$ within a band $0< y < y_v$, where $y_v =W(1-H_{stop}/H_z$).  From the London equation  and the fact that a vortex contributes no net current along the sample length,  it follows that this band of vortices generates a current density 
\begin{eqnarray}
K_{vx}(y)\!&=&\!(H_z/W\Lambda)[2W(y_v\!-\!y)\!-\!y_v^2], 
\;0 \le y \le y_v,\\
&=&\!-H_z y_v^2/W\Lambda,\;\;\;\;\;\;\;\;\;\;\;\;\;\;\;\;\;\;\;\;\;\;y_v \le y \le W.
\end{eqnarray}
The net sheet-current density in general is
\begin{equation}
K_x(y) = K_I + K_{Hx}(y) + K_{vx}(y),
\label{Kxwithvortices}
\end{equation}
but since we must also  have $K_x(y) = 0$ within the vortex-filled region ($0< y <y_v$) so that these vortices do not move, we obtain
\begin{equation}
K_I = (H_z/W\Lambda)(W-y_2)^2.
\label{KIcondition}
\end{equation}

At the critical current,  the vortices present in the region $0 < y < y_2$ produce the additional sheet current $K_{vx}(W) = -H_z y_v^2/W\Lambda$.  To determine the condition for which a new vortex is nucleated using Eq.\ (\ref{Gsmalldelta}), we must therefore add to the right-hand side an additional term $\phi_0H_z y_v^2\delta/W\Lambda$.  The quantity in parentheses in the denominator of Eq.\ (\ref{deltab}) is then replaced by $K_I + H_zW/\Lambda-H_zy_v^2/W\Lambda$. When this quantity reaches the value $K_{cs}$, $K_I$ reaches its critical value, $K_{cv}(H_z)$, such that 
\begin{equation}
K_I=K_{cs}- H_zW/\Lambda+H_zy_v^2/W\Lambda.
\label{Kentrycondition}
\end{equation}  
 Setting $K_I = K_{cv}(H_z)$ in Eqs.\ (\ref{KIcondition}) and (\ref{Kentrycondition}), we obtain for $H_z \ge H_{\rm stop}$,
\begin{eqnarray}
y_2 &=& W(1-H_{\rm stop}/H_z), \\
K_{cv}(H_z)&=&K_c(0)(H_{\rm stop}/2H_z).
\end{eqnarray}
Thus, in this approximation, the linear behavior $K_{cv}(H_z)= K_c(0)(1-H_z/2H_{\rm stop})$ for $0 < H_z < H_{\rm stop}$ turns into a slower $K_{cv}(H_z)=K_c(0)(H_{\rm stop}/2H_z)$ decrease for $H_z > H_{\rm stop}$ with no change in slope at $H_z = H_{\rm stop}$, a behavior similar to that found in Ref.\ \onlinecite{Benkraouda98}. 

Note that the calculations given here assume that the sheet-current density in the strip is much higher than the critical sheet current attributable to bulk pinning of vortices or antivortices.  The effects of bulk pinning could be accounted for using a treatment similar to that in Ref.\ \onlinecite{Elistratov02}.

The solid curves in Fig.\ \ref{Kcv&KcaFig} show the critical sheet current of a long straight strip (normalized to $K_{cs}$) vs $H_z$.  For $H_z > 0$, the critical sheet current is $K_{cv}(H_z)$, which is determined by the nucleation of vortices (hence the subscript v) at $y = W$, while for $H_z < 0$, the critical sheet current is $K_{ca}(H_z)$, which is determined by the nucleation of antivortices (hence the subscript a) at $y = 0$.  The dashed curves show the extension of the linear portions of the curves.  The mirror symmetry about the vertical axis at $H_z$ occurs because of the mirror symmetry of the sample geometry about $y = W/2$.  As we show in the next section, if the sample geometry is asymmetric, there is a dramatic difference in the behavior of the critical sheet current depending upon the direction of the applied field.

\section{$I_c(H_z)$ for strips with left turns
\label{Turns}} 

\begin{figure}
\includegraphics[width=6cm]{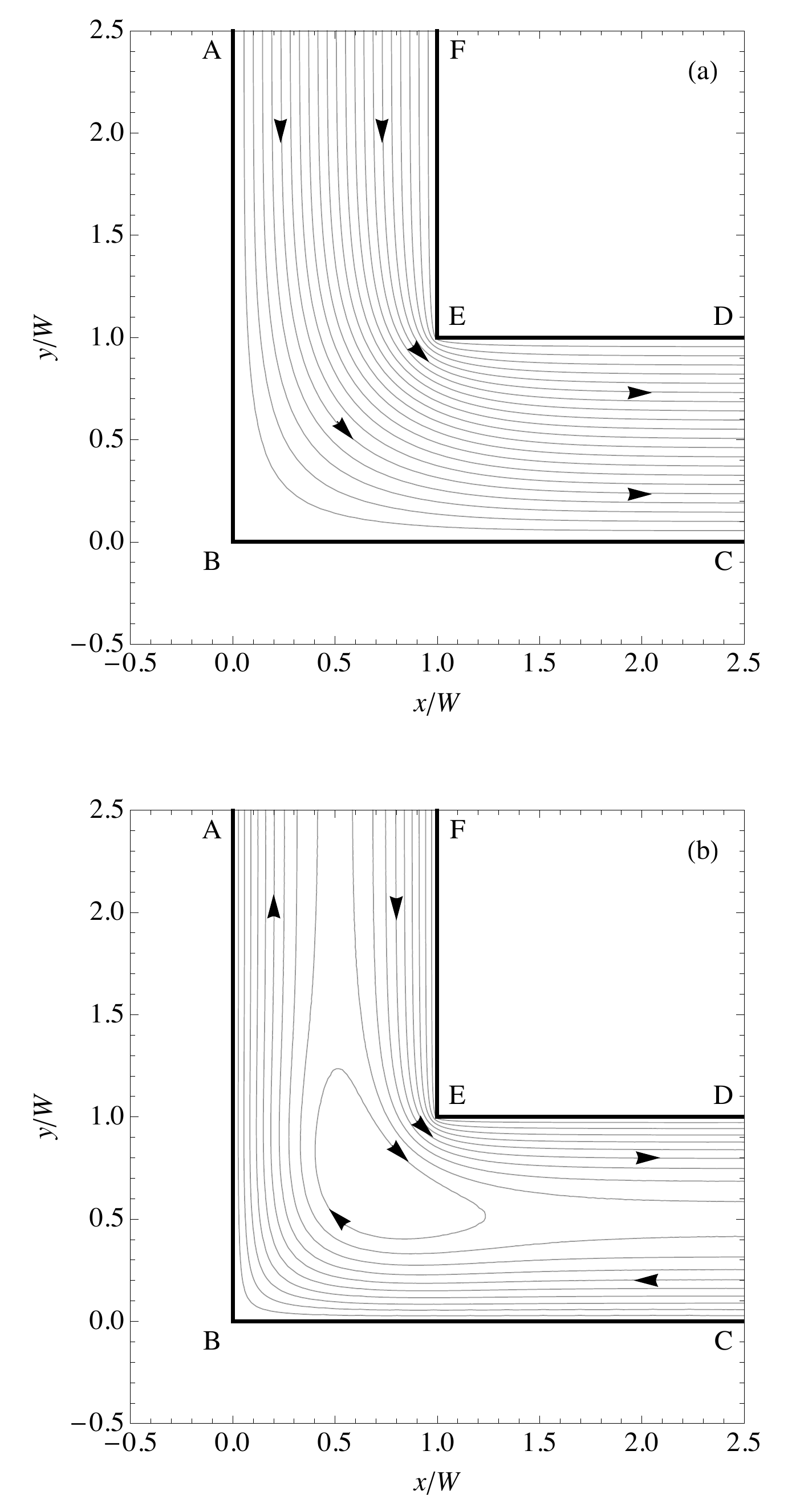}
\caption{%
(a) Current flow in a strip carrying current $K_I W$ around a left-hand 90$^\circ$ turn, shown by the contour plot of its stream function $S_{Iz}(x,y)$.  The contours correspond to streamlines of the induced sheet-current density $\bm K_I(x,y)$, and the arrows show the direction of the current.  Note the current crowding that occurs near the inner corner E.
(b) Clockwise current flow induced by a positive applied field $H_z$, shown by the contour plot of the stream function $S_{Hz}(x,y)$.  The contours correspond to streamlines of the field-induced sheet-current density $\bm K_H(x,y)$, and the arrows show the direction of the current.  Note the current crowding that occurs near the inner corner E.}
\label{90Fig}
\end{figure}

\begin{figure}
\includegraphics[width=6cm]{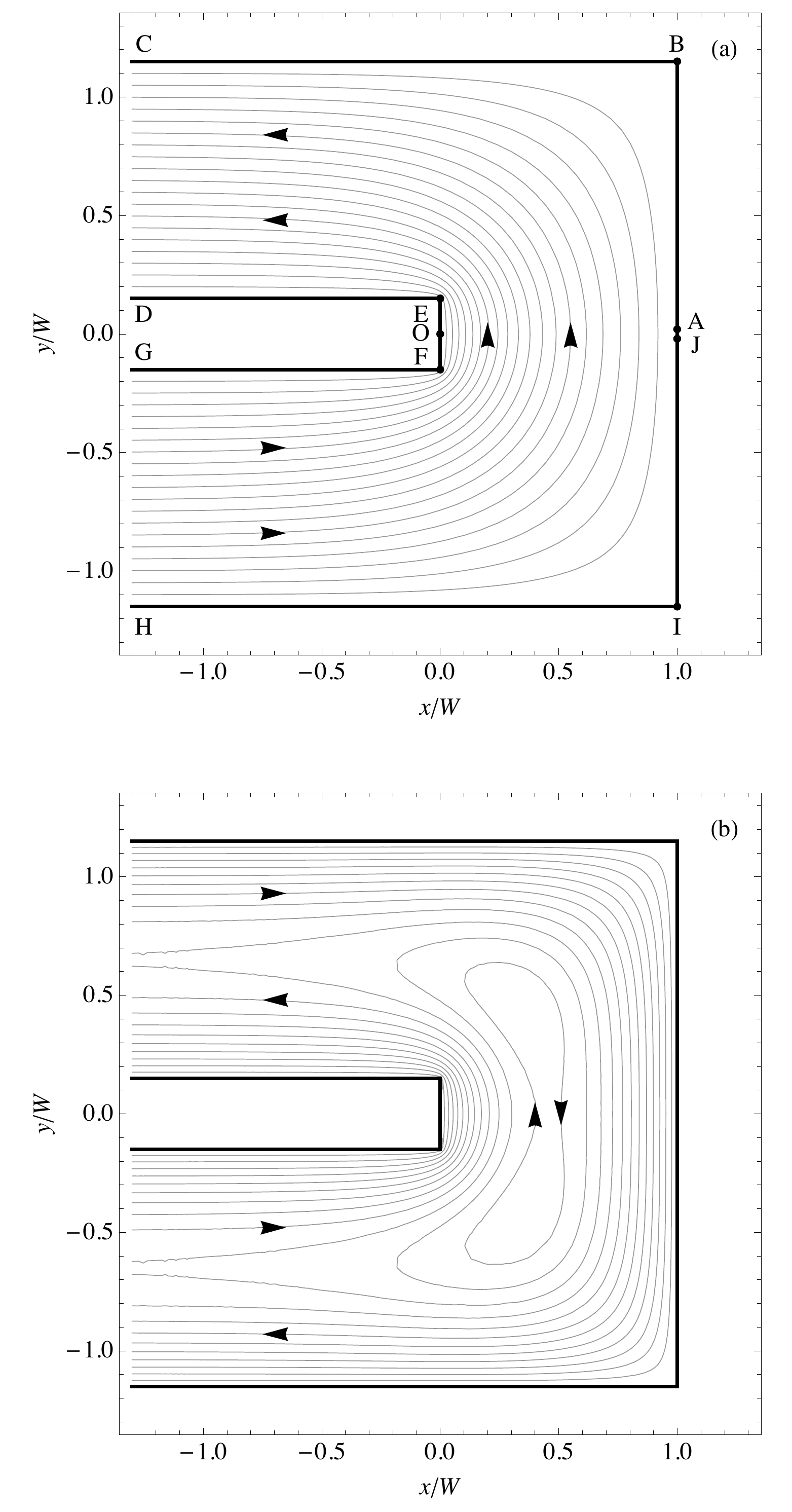}
\caption{%
(a) Current flow in a strip carrying current $K_I W$ around a sharp rectangular 180$^\circ$ turn, shown by the contour plot of its stream function $S_{Iz}(x,y)$.  The contours correspond to streamlines of the induced sheet-current density $\bm K_I(x,y)$, and the arrows show the direction of the current.  Note the current crowding that occurs near the inner corners E and F.
(b) Clockwise current flow induced by a positive applied field $H_z$, shown by the contour plot of the stream function $S_{Hz}(x,y)$.  The contours correspond to streamlines of the field-induced sheet-current density $\bm K_H(x,y)$, and the arrows show the direction of the current.  Note the current crowding that occurs near the inner corners.}
\label{rect180Fig}
\end{figure}

\begin{figure}
\includegraphics[width=6cm]{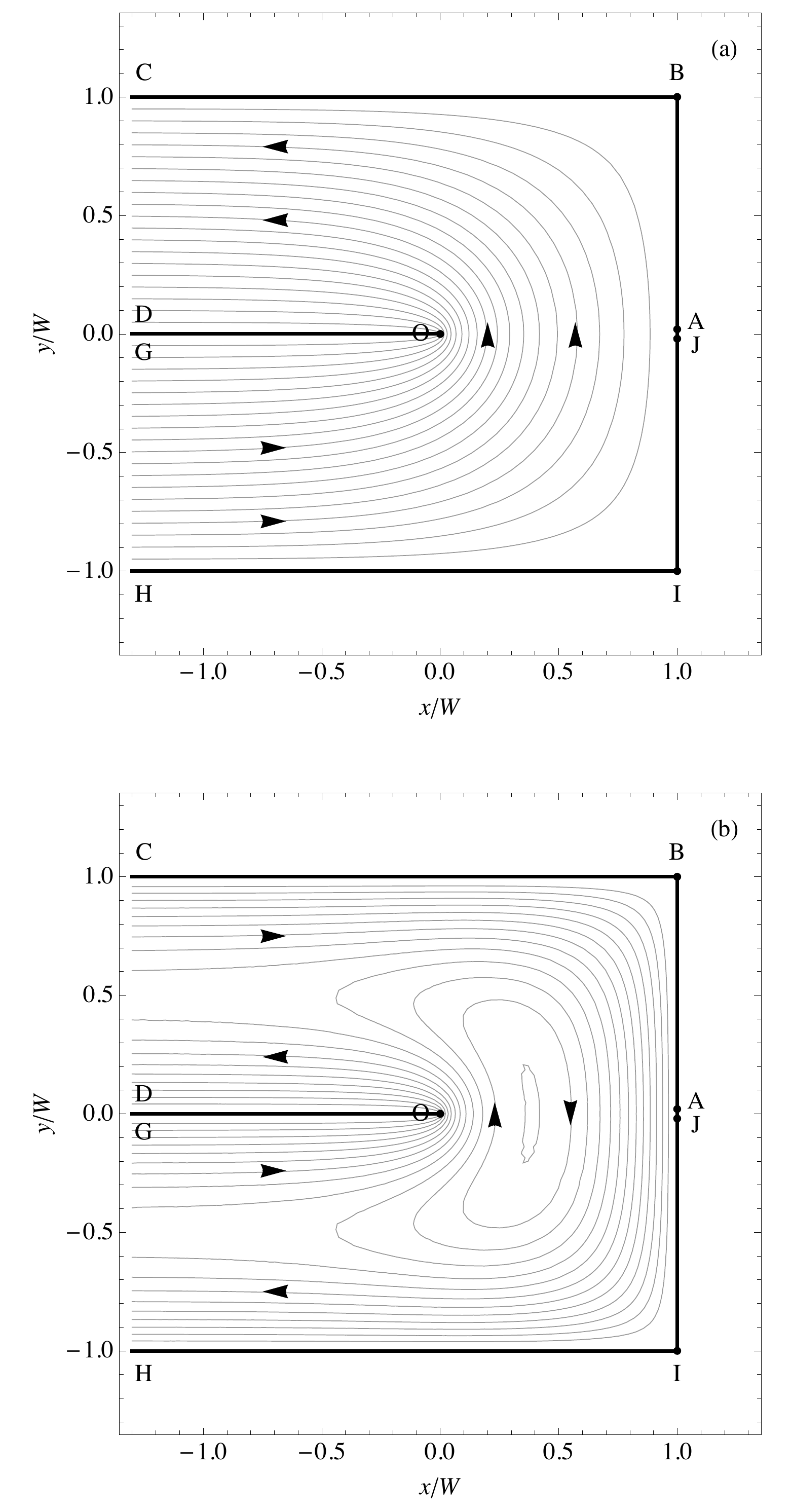}
\caption{%
(a) Current flow in a strip carrying current $K_I W$ around a sharp  180$^\circ$ turn, shown by the contour plot of its stream function $S_{Iz}(x,y)$.  The contours correspond to streamlines of the induced sheet-current density $\bm K_I(x,y)$, and the arrows show the direction of the current.  Note the current crowding that occurs near the origin O.
(b) Clockwise current flow induced by a positive applied field $H_z$, shown by the contour plot of the stream function $S_{Hz}(x,y)$.  The contours correspond to streamlines of the field-induced sheet-current density $\bm K_H(x,y)$, and the arrows show the direction of the current.  Note the current crowding that occurs near the origin.}
\label{sharp180Fig}
\end{figure}
Next we consider strips of width $W$ carrying current $K_I W$ making left turns around a sharp 90-degree left turn [Fig.\ \ref{90Fig}(a)],  a sharp rectangular 180-degree turnaround [Fig.\ \ref{rect180Fig}(a)], or a sharp 180-degree turnaround [Fig.\ \ref{sharp180Fig}(a)].  In Ref.\ \onlinecite{Clem11},  using a London-model description of vortices and antivortices, Clem and Berggren developed a procedure for estimating  how the critical sheet-current density in the absence of an applied field $K_c(0)$ is reduced because of current crowding at the inner corners of turns and bends.  For each sample geometry considered, they carried out a calculation of the Gibbs free energy similar to that in Eq.\ (\ref{Gsmalldelta}) and showed that when $\xi \ll W$, the critical sheet-current density can be expressed as $K_c(0) = K_{cs} R$, where $K_{cs}$ is the critical sheet current for a long straight strip in the absence of an applied field [Eq.\ (\ref{Kcstrip0})]  and $R$
is a critical-current reduction factor $(0<R < 1)$.  For left-hand turns, we can append the subscript $v$ to the critical sheet-current density $K_{cv}(0)$ as a reminder that the critical current occurs at the threshold for the nucleation of vortices at the sharp inner corners.  Antivortices could be nucleated along the long, straight portions of the outer boundaries only at a higher current density, $K_{ca}(0) = K_{cs}$.  

In this section, we extend the approach of Ref.\ \onlinecite{Clem11} to study the effect of an applied perpendicular magnetic field $\bm H = \hat z H_z$ upon the critical sheet-current density for the sample geometries shown in Figs.\ \ref{90Fig}-\ref{sharp180Fig}.  When the applied current makes only left turns, as shown in 
Figs.\ \ref{90Fig}(a),  \ref{rect180Fig}(a), and \ref{sharp180Fig}(a), the critical current for small $H_z$ is dominated by the  nucleation of vortices at the sharp inner corners, and we therefore seek expressions for $K_{cv}(H_z)$, where the subscript $v$ denotes vortices, which carry magnetic flux in the $+z$ direction.  The critical-current calculation requires knowledge of the detailed form (especially the behavior of the divergences near the inside corners) of the applied sheet-current density   $\bm K_I(x,y) = \nabla \times \bm S_I(x,y)$, where $\bm S_I(x,y) = \hat z S_{Iz}(x,y)$ is its stream function, and the magnetic-field-induced sheet-current density $\bm K_H = \nabla \times \bm S_H$, where $\bm S_H = \hat z S_{Hz}$ is its stream function.   
Figures \ref{90Fig}(a),  \ref{rect180Fig}(a), and \ref{sharp180Fig}(a) show contour plots of the stream functions $S_I(x,y)$, and Figs.\ \ref{90Fig}(b),  \ref{rect180Fig}(b), and \ref{sharp180Fig}(b) show contour plots of the stream functions $S_H(x,y)$.   For each of the three geometries, the field-dependent critical sheet-current densities for the nucleation of vortices at the inner corners can be expressed as
\begin{equation}
K_{cv}(H_z) = K_{cv}(0)-\sigma WH_z/\Lambda,
\label{Kcv}
\end{equation}
where $K_{cv}(0) = K_{cs} R$. The specific forms of the critical-current-reduction factor $R$ and the field-slope parameter $\sigma$ ($0 <  \sigma < 1$) depend upon the sample geometry. The details of the calculations of $S_I(x,y)$, $S_H(x,y)$, $R$, $\sigma$, and $K_{cv}(H_z)$ are given in the Appendix. 

Note that $K_{cv}$ decreases with increasing $H_z$, because the field-induced sheet-current density $\bm K_H$  is in the {\it same} direction as the applied sheet-current density at the vortex-nucleation site.  $K_{cv}$ increases for negative $H_z$ of increasing magnitude, because the  field-induced sheet-current density $\bm K_H$  is then in the {\it opposite} direction as the applied sheet-current density at the vortex-nucleation site. 

We next examine how the application of a magnetic field $\bm H = \hat z H_z$ affects the critical current due to the nucleation of antivortices at the outer boundaries of the structures shown in Figs.\ \ref{90Fig}-\ref{sharp180Fig}.  This occurs where the magnitude of the sheet-current density $\bm K(x,y) = \bm K_I(x,y) + \bm K_H(x,y)$ has its maximum value.   It can be shown that, except for negative values of $H_z$ of large magnitude, this maximum occurs on the outer boundary at large distances from the outer corners, where the spatial dependence of $\bm K(x,y) = \bm K_I(x,y) + \bm K_H(x,y)$ becomes the same as in a long, straight strip of width $W$. Thus, the critical sheet-current density at which antivortex nucleation occurs is 
\begin{equation}
K_{ca}(H_z) = K_{cs}+WH_z/\Lambda.
\label{Kca}
\end{equation}
  The critical current for the nucleation of antivortices increases linearly from the value $K_{cs}$ for $H_z>0$ and decreases linearly for $H_z < 0$, just as in the case of an infinitely long strip, as given in Eq.\ (\ref{KcstripH}) and  shown in Fig.\ \ref{Kcv&KcaFig} by the curve $K_{ca}(H_z)$.

\begin{figure}
\includegraphics[width=8cm]{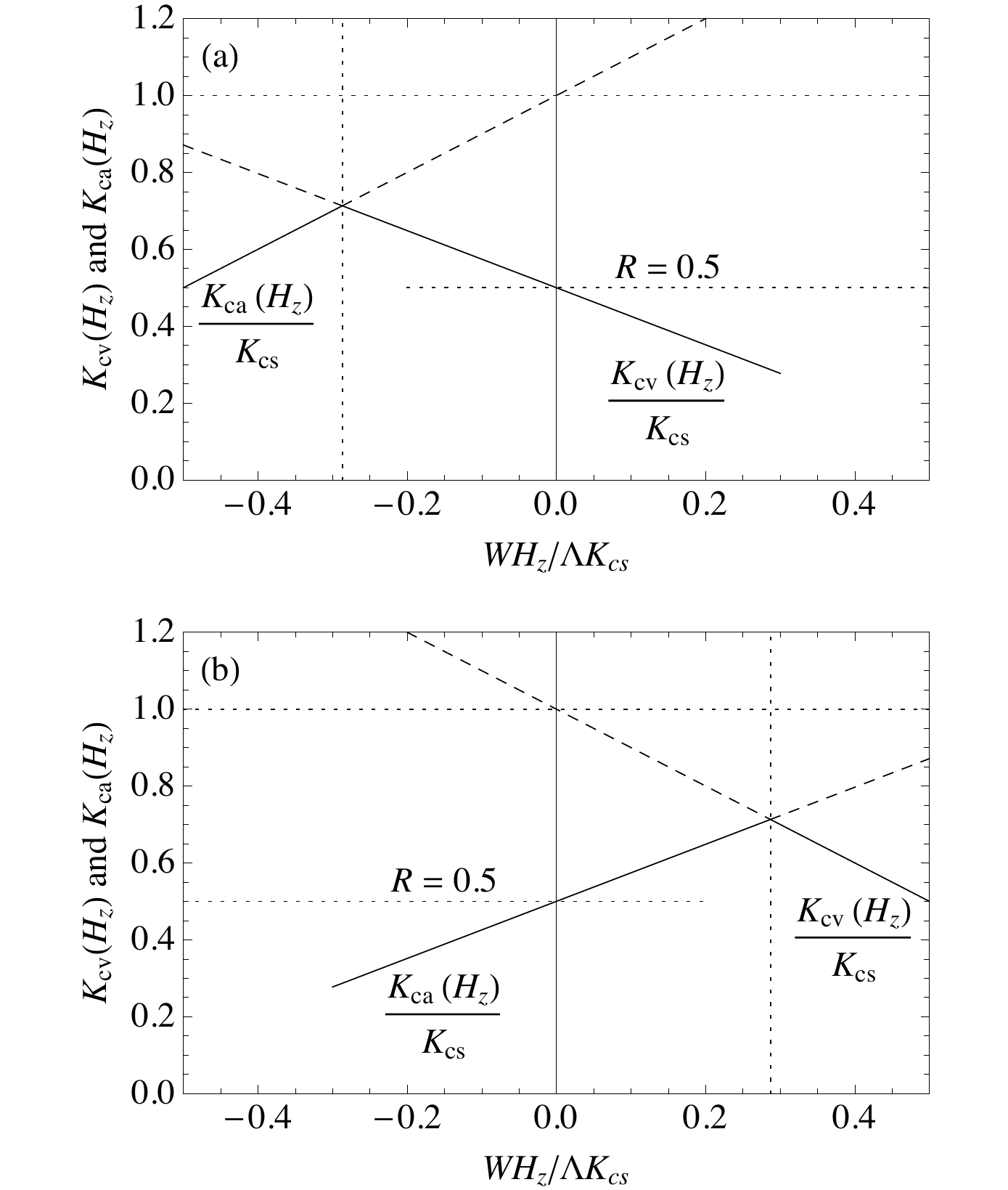}
\caption{%
(a) Left turns:  Generic behavior of the critical sheet current for vortices $K_{cv}(H_z)$ and antivortices $K_{ca}(H_a)$ (normalized to $K_{cs}$) vs $H_z$  (normalized to $\Lambda K_{cs}/W$), illustrated here for a sharp 90-degree turn when the current $I$ flows down the vertical portion of the strip shown in Fig.\ \ref{90Fig}(a) and makes a left turn around the corner E.  (b) Right turns:  Same as (a) except for the case that the current $I$ flows in through the horizontal portion of the strip shown in Fig.\ \ref{90Fig}(a) and makes a right turn around the corner E.  The values $\xi/W = 0.047$ and $R = 0.5$ were assumed for both plots, such that $WH_{p}/\Lambda K_{cs} = 0.287$ and $K_{cp}/K_{cs} = 0.713$.   }
\label{Kcv&KcaFig90}
\end{figure}

The solid lines in Fig.\ \ref{Kcv&KcaFig90}(a) show the generic behavior of the critical sheet-current density for all three left-hand turns shown in Figs.\ \ref{90Fig}-\ref{sharp180Fig}, but illustrated using parameters for the sharp 90-degree turn, $R = 0.5$ and $\sigma = 8\bm G/\pi^2 = 0.742,$ where $\bm G = 0.915965...$ is Catalan's constant.  The overall critical current of films with sharp left-hand turns is determined by the critical current for the nucleation of vortices for all values of $H_z > H_{max}$ and by the critical current for the nucleation of antivortices for all values of $H_z < H_{max}$, where $H_{max}$ is the applied field at which the critical currents for vortices and antivortices are equal.  From Eqs.\ (\ref{Kcv}) and (\ref{Kca}) 
we see that 
when the applied field is $H_{max} = - H_p$, where
\begin{equation}
H_p = \Big(\frac{\Lambda K_{cs}}{W}\Big)\frac{(1-R)}{(1+\sigma)}=
\frac{2\xi(1-R)}{eW(1+\sigma)}H_{c2},
\label{Hp}
\end{equation}
the maximum overall critical sheet current has the peak value  $K_c(H_{max})=K_{cp}$, where
\begin{equation}
\frac{K_{cp}}{K_{cv}(0)} =\frac{(R+\sigma)}{R(1+\sigma)}.
\label{Kcp}
\end{equation}
The latter equality in Eq.\ (\ref{Hp}) arises from Eq.\ (\ref{Kcstrip0}) and the well-known expression $\mu_0 H_{c2} =\phi_0/2\pi\xi^2$.   
In summary, we see that by applying a negative applied field $H_z =H_{max} =  -H_p$, the critical current, which in zero field is suppressed by current crowding at the sharp inner corner, can be increased by the factor $(R+\sigma)/R(1+\sigma)$.  This factor is very large when $R$ is small, is equal to 1.43 for the sharp 90-degree turn when $R = 0.5$ and $\sigma =  0.742,$ as shown in Fig.\ \ref{Kcv&KcaFig90}, and approaches 1 as $R \to 1$.  Expressions for $R$ and $\sigma$ for all three geometries shown in Figs.\ \ref{90Fig}-\ref{sharp180Fig} are given in the Appendix.  

The generic behavior of the field-dependent critical current when the current makes {\it right} turns is shown in Fig.\ \ref{Kcv&KcaFig90}(b).   Consider what happens in a strip with a sharp 90-degree turn as in Fig.\  \ref{90Fig}(a) when the direction of the current $I$ is reversed.  
 At $H_z=0$, the critical current is now determined by the onset of {\it antivortex} nucleation at E.  This critical current increases linearly with $H_z$, but the maximum overall critical current is reached when  $H_{max}=H_p$ and the critical current for antivortex nucleation at E becomes equal to the critical current for vortex nucleation along one of the straight sides far from the outer corner B. The overall critical sheet current is shown by the solid lines in Fig.\ \ref{Kcv&KcaFig90}(b).

The above analysis, which assumes $\xi \ll W$, is intended to describe the critical current only at relatively low fields, where the behavior is linear in the applied field.  At high applied fields, vortices or antivortices are forced into the strip, and the dependence of the critical current becomes nonlinear, similar to the behavior discussed in Sec.\ \ref{nonlinear}.  The effect of the vortex distribution inside the sample on the magnetic-field dependence of the critical current will be addressed in the next section.

\section{Numerical simulations within the time-dependent Ginzburg-Landau theory\label{TDGL}}

\begin{figure}[b]
\includegraphics[width=\linewidth]{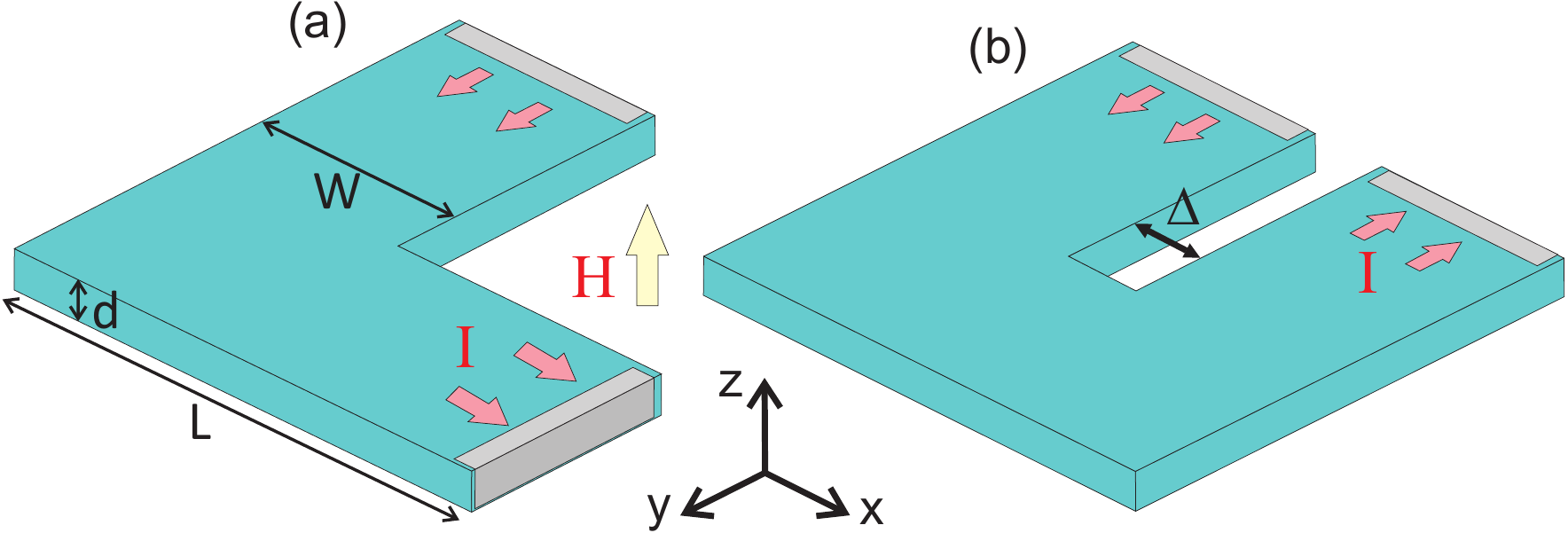}
\caption{\label{fig1}(Color online) The model systems: Superconducting strips (of length $L$, width $W$ and thickness $d$) with (a) 90$^{\circ}$ and (b) 180$^{\circ}$ turns. The current is applied through normal-metal contacts (and flows always such that it makes a left turn) and the output voltage is measured at a small distance away from these leads. The magnetic field is applied perpendicular to the sample either in the $z$- (positive field) or the -$z$-direction (negative field).}
\end{figure}

In order to confirm the theoretical predictions in the preceding sections, we performed numerical simulations within the time-dependent Ginzburg-Landau (GL) theory. We consider superconducting strips (with thickness $d\ll\lambda$ and width $w\ll\Lambda=2\lambda/d^2$) with multiple sharp turns in the presence of a transport current (applied through the normal contacts) and a perpendicular magnetic field of magnitude $H$ (see Fig.\  \ref{fig1}).

\subsection{Theoretical approach}

To study the dynamics of the superconducting condensate in such complex structures, we used the following generalized time-dependent Ginzburg-Landau (GL) equation \cite{Kramer}:
\begin{eqnarray}
\frac{u}{\sqrt{1+\gamma^2|\psi|^2}}\left(\frac{\partial}{\partial t}\!+\!i\varphi\!+\!\frac{\gamma^2}{2}\frac{\partial |\psi|^2}{\partial t} \right)\! \psi =(\nabla-i\mathbf{A})^2\psi\nonumber\\
+(1-|\psi|^2)\psi,
\end{eqnarray}
which is coupled with the equation for the electrostatic potential $\Delta\varphi={\rm div}\{\textrm{Im}[\psi^*(\nabla-{\rm i}{\bf A})\psi]\}$.
Here distance is scaled to $\xi$, the vector potential ${\bf A}$ is in units of $\phi_0/2\pi\xi$, time is in units of the GL relaxation time $t_{GL}=\mu_0\lambda^2/\rho_n$ ($\rho_n$ is the normal-state resistivity), and voltage is scaled to $V_0=\phi_0/2\pi t_{GL}$. The coefficient $u$, which governs the relaxation of the order parameter (i.e., the ratio between relaxation times for the phase and the amplitude of $\psi$), and the material parameter $\gamma$ are chosen as $u=5.79$ and $\gamma=10$, which are found within the microscopic BCS theory for superconductors with weak depairing. \cite{Kramer} Using the normal-state resistivity $\rho_n=18.7$ $\mu\Omega$cm, zero-temperature coherence length $\xi(0)=10$ nm and penetration depth $\lambda(0)=200$ nm, which are typical for Nb thin films, \cite{Gubin} one can obtain $t_{GL}\approx2.69$ ps and $V_0\approx0.12$ mV near $T_c$. We use superconducting-vacuum boundary conditions $(\nabla-{\rm i}{\bf A})\psi|_n=0$ and $\nabla\varphi|_n=0$ at all sample boundaries, except at the current contacts where we use $\psi=0$ and $\nabla\varphi|_n=-j$, with $j$ being the applied current density in units of $j_0=\phi_0/2\pi\mu_0\xi\lambda^2$. Assuming that $W\ll\Lambda=2\lambda/d^2$ we neglect  demagnetization effects and chose ${\bf A}=\mu_0(-Hy/2,Hx/2)$. We solve the above coupled non-linear differential equations self-consistently in 2D using Euler [for Eq.\ (19)] and multi-grid \cite{multigrid} (for the equation of the electrostatic potential) iterative procedures.

\begin{figure} [b]
\includegraphics[width=\linewidth]{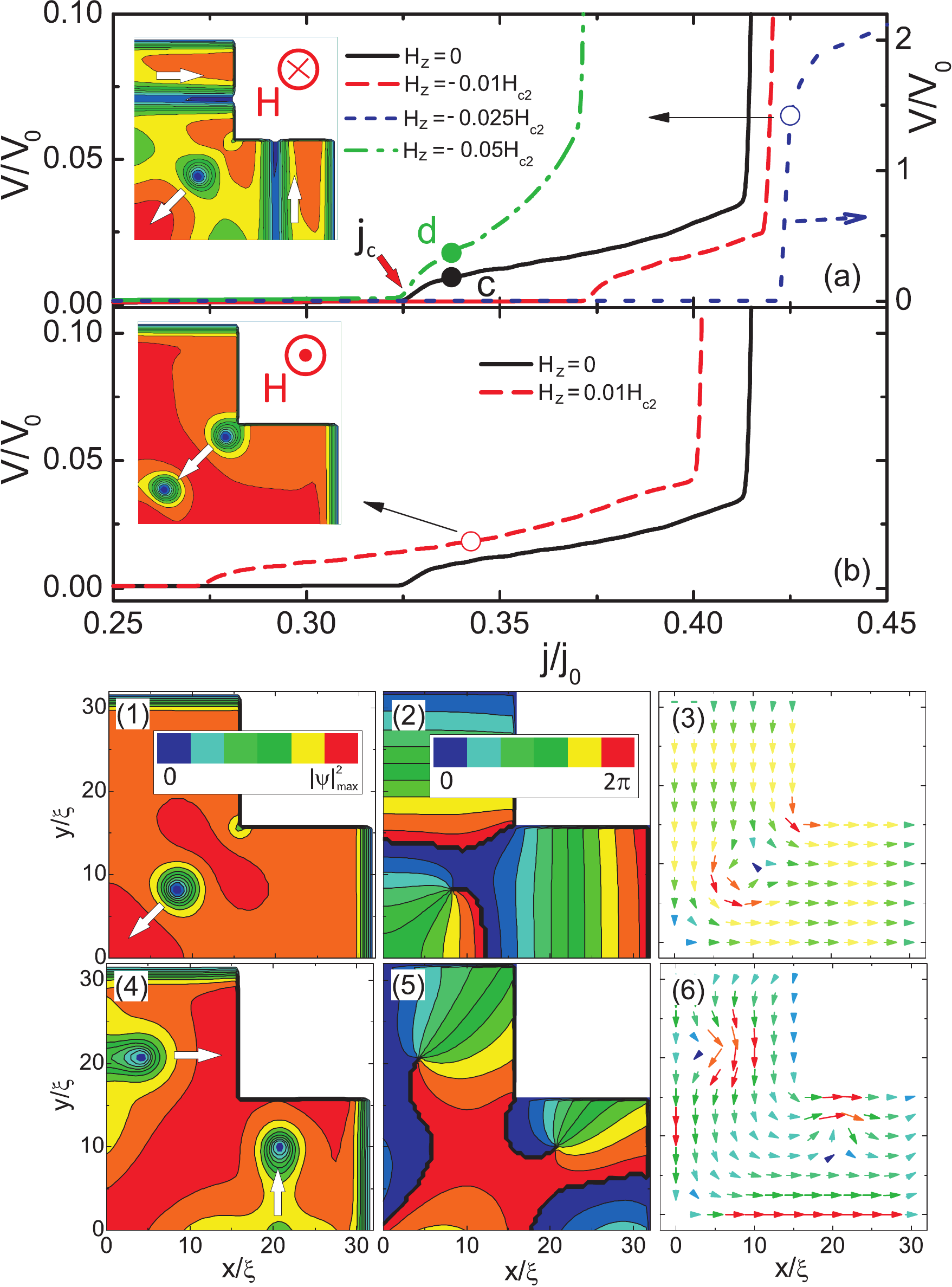}
\caption{\label{fig2}(Color online) Current-voltage characteristics of the sample with a 90$^{\circ}$ turn for different values of (a) negative and  (b) positive magnetic field. Right axis in (a) shows the voltage values for $H_z=-0.025H_{c2}$. Dimensions of the sample are $L=32\xi$ and $W=15.5\xi$. Insets show snapshots of the Cooper-pair density at the current and magnetic field values indicated by open circles in the $I-V$ curves. Panels 1-6 show snapshots of the Cooper-pair density (1, 4), the phase of the order parameter (2, 5), and the supercurrent distribution (3, 6) for vortex nucleation at zero magnetic field (1-3) [point ``c'' in (a)] and  antivortex nucleation at $H_z= -0.05H_{c2}$ (4-6) [point``d'' in (a)]. White arrows indicate the direction of vortex or antivortex motion.}
\end{figure}

\subsection{Superconducting strip with a 90-degree turn}

We begin our analysis by demonstrating the properties of a superconducting strip with a 90$^{\circ}$ turn [see Fig.\  \ref{fig1}(a)] by constructing the time-averaged voltage vs applied current ($I-V$) and the voltage vs time $V(t)$ characteristics of the sample for a constant magnetic field. The time-averaged voltage shown in Figs. 7, 11, and 13 is equivalent to $V = \phi_0 \nu$, where $\nu$ is the net rate with which vortices cross the strip in one direction and antivortices cross the strip in the opposite direction, in accord with the Josephson relation.\cite{Josephson62,Josephson65}As a representative example, we consider a superconducting strip with length $L=32\xi$ and width $W=15.5\xi$, the $I-V$ characteristics of which are shown in Fig.\  \ref{fig2} for different values of negative (a) and positive (b) magnetic field. We first discuss the results for negative direction of the magnetic field, which is the situation when the screening (Meissner) currents oppose the applied current near the sharp inner corner of the sample. In the absence of the magnetic field [solid black curve in Fig.\  \ref{fig2}(a)] zero resistance of the sample is maintained up to a threshold current density $j_{c}=0.3275j_0$, above which the system goes into the resistive state with a finite-voltage jump. This resistive state is characterized by the periodic nucleation of vortices near the inner corner where the current density is highest (see panels 1-3 in Fig.\  \ref{fig2}). This vortex is driven further by the Lorentz force towards the outer corner of the sample where it leaves the sample (see panel 1). The nucleation rate of vortices in the inner corner increases with further increasing the applied current, and at sufficiently large currents the system transits to a higher dissipative state, characterized by fast-moving (kinematic) vortices. \cite{Andronov,golib} The critical current of the sample $j_{c}$ considerably increases with applying small negative magnetic field [dashed red curve in Fig.\  \ref{fig2}(a)]. This is because the Meissner currents reduce the current crowding at the sharp inner corner. However, at larger values of the negative field the critical current becomes smaller [see dot-dashed green curves in Fig.\  \ref{fig2}(a)], because antivortices start penetrating the sample (see panels 4-6 in Fig.\  \ref{fig2}). These antivortices (compare phase plots in panels 2 and 5) nucleate away from the corners of the sample (where the current density is maximal) and are driven across the sample (white arrows indicate the direction of motion). Thus, the critical current of the sample has a non-monotonic dependence on the negative magnetic field. At intermediate values of the magnetic field, we observed the coexistence of vortices and (kinematic) antivortices as shown in the inset of Fig.\  \ref{fig2}(a) (see also discussion of Fig.\  \ref{fig5}). 

The dashed red curve in Fig.\ \ref{fig2}(b) shows the $I-V$ curve of the sample for positive magnetic field. For this direction of the field the Meissner currents add to the applied current near the inner corner of the sample, reducing the surface barrier for the nucleation of a vortex there.
Indeed, analysis of the temporal characteristics of the sample (not shown here) shows that the vortices always nucleate near the inner sharp corners, as illustrated in the inset of Fig.\  \ref{fig2}(b). Thus, the critical current is solely determined by the vortex entry at the inner corner, and it becomes a monotonically decreasing function of the magnetic field.

\begin{figure}[t]
\includegraphics[width=\linewidth]{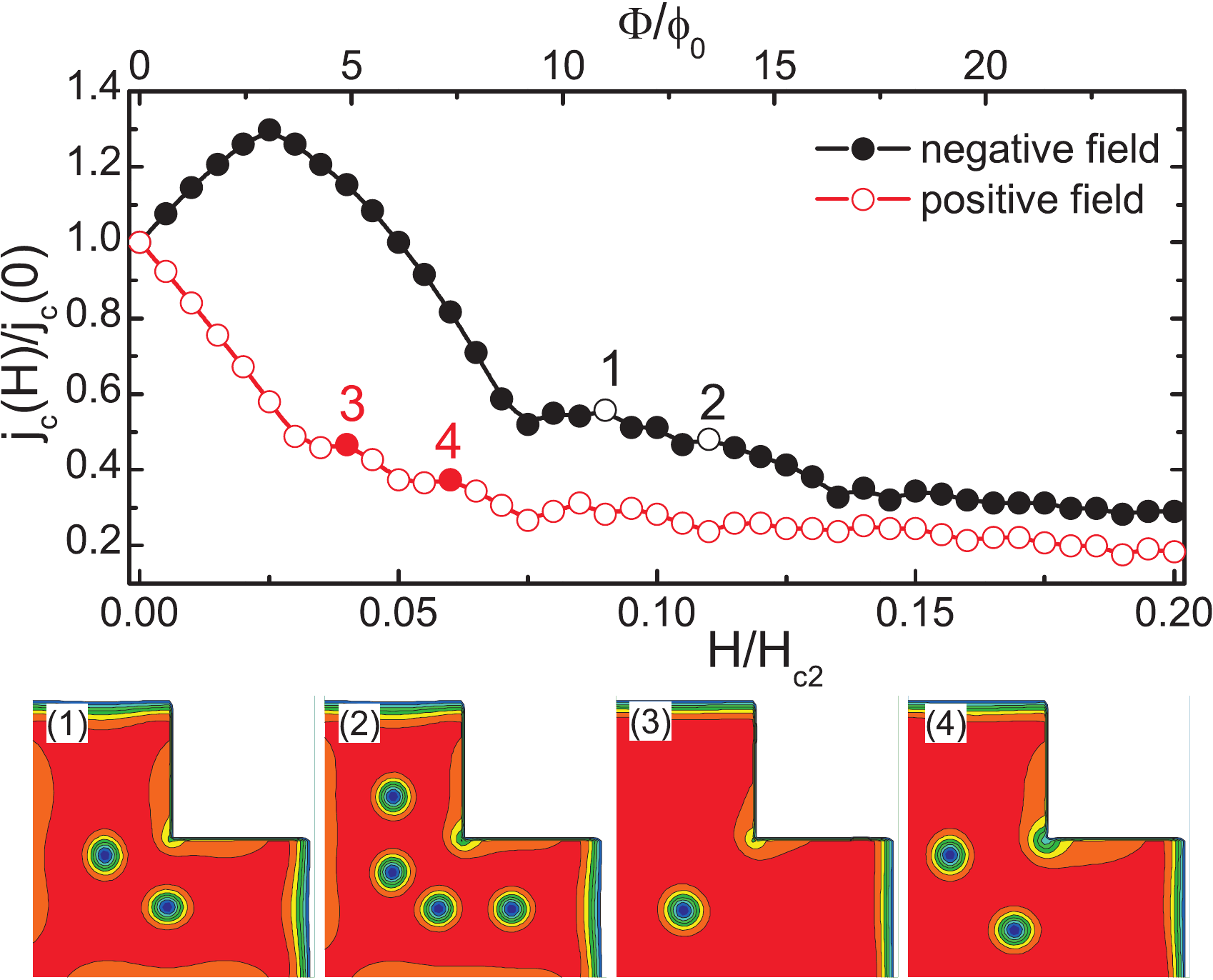}
\caption{\label{fig3}(Color online) The critical current $j_{c}$ of the sample in Fig.\  \ref{fig2} as a function of negative (filled black circles) and positive (open red circles) applied magnetic field. The results are normalized to the critical current at zero magnetic field. Top axis shows the flux (in units of the flux quantum $\phi_0$) through the sample. Panels 1-4 show contour plots of $|\psi|^2$ for magnetic field values indicated on the $j_{c}(H)$ curves and for current values just below $j_{c}$.}
\end{figure}

\begin{figure} 
\includegraphics[width=\linewidth]{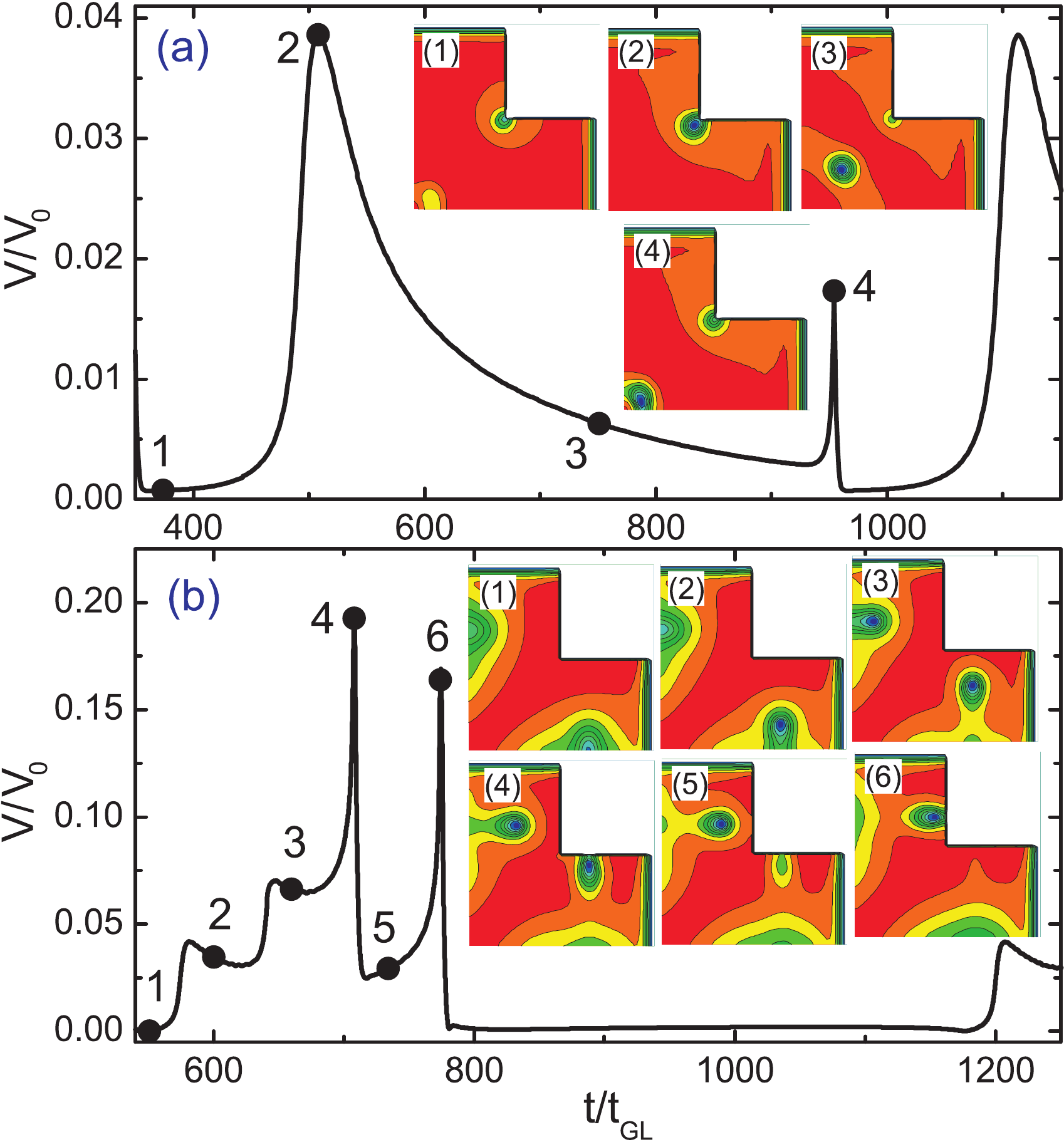}
\caption{\label{fig4}(Color online) Voltage vs time response of the sample in Fig.\  \ref{fig2} at $j=0.34j_0$ for (a) $H_z=0$ and (b) $H_z=-0.05H_{c2}$, upon transition to the resistive state [see points ``c'' and ``d'' in Fig.\  \ref{fig2}(a)]. Insets show snapshots of the Cooper-pair density at the times indicated on the $V(t)$ curves, which illustrate the motion of vortices (a) and antivortices (b).}
\end{figure}

\begin{figure}[b]
\includegraphics[width=\linewidth]{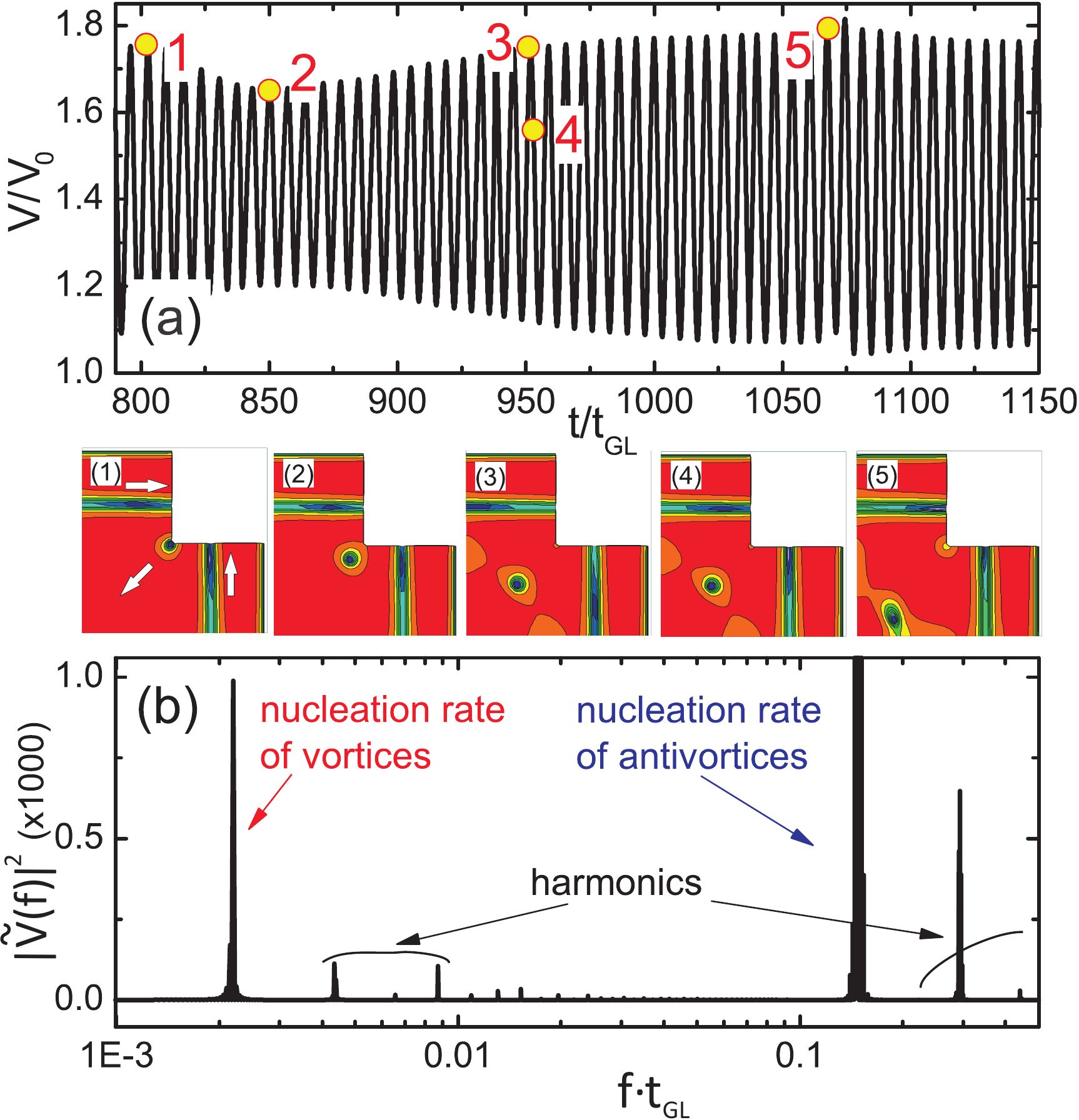}
\caption{\label{fig5}(Color online) (a) Time evolution of the voltage of the sample in Fig.\  \ref{fig2} for $j=0.425j_0$ and $H_z=-0.025H_{c2}$. Panels 1-5 show the time evolution of the superconducting  condensate at the time intervals indicated on the $V(t)$ curve. (b) The Fourier power spectrum of $V(t)$. The nucleation rates of vortices and antivortices are indicated.}
\end{figure}

Figure \ref{fig3} summarizes our findings, where we plot the resistive-state transition current $j_{c}$ as a function of applied magnetic field (see top axis for the flux going through the sample). At relatively low negative fields (filled black circles), the critical current increases with increasing magnetic field; up to 30\% enhancement can be achieved for the given parameters of the sample. However for high magnetic fields, $j_{c}$ decreases again with increasing negative field, because magnetic-field-induced antivortices start to nucleate [see the discussion of Fig.\  \ref{fig4}(b)]. At larger fields, the critical current shows a diffraction-like pattern as a function of magnetic field. This is because different antivortex states are stabilized before the system transits to the resistive state (see panels 1 and 2). For the positive direction of the magnetic field (open red circles), the critical current is a monotonically decreasing function of  $H_z$: at relatively low fields $j_{c}$ has a linear dependence on the field. At higher fields, vortices start penetrating the sample (see panels 3 and 4) and the $j_{c}(H)$ curve becomes nonlinear -- the behavior found for straight superconducting strips (see the discussion in Sec. IIB and also Ref. \onlinecite{Elistratov02}). Notice that the critical current for negative magnetic field is always larger than the one for positive field due to the suppression of current crowding at the sharp inner corner.

To get a better insight into the dynamics of vortices in the system, we plotted in Fig.\  \ref{fig4}(a) the time evolution of the output voltage at zero magnetic field and for an applied current just above $j_{c}$, together with snapshots of the Cooper-pair density at times indicated in the $V(t)$ curves. The output voltage oscillates periodically in time (one full period is shown) with a minimum corresponding to the Meissner state [inset 1 in Fig.\  \ref{fig4}(a)]. This voltage can be described as  a periodic sequence of single-vortex pulses, each of integrated area $\int V dt = \phi_0$.\cite{Clem70,Clem81}  With time, a vortex nucleates at the inner corner of the sample (inset 2) where the current density is higher due to the current crowding. This nucleation process corresponds to a maximum in the $V(t)$ curve. By means of the Lorentz force this vortex moves deeper inside the sample towards the outer corner of the sample (inset 3). Note that the voltage decreases during the motion of the vortex, indicating that the vortex slows down during its motion. The expulsion of the vortex also leads to an extra voltage peak (inset 4). Note that for this value of the current there is only one vortex in the sample at a given time; however, for larger currents more than one vortex can be present [see the inset in Fig.\  \ref{fig2}(b)].

Figure \ref{fig4}(b) shows the voltage characteristics of the sample together with the evolution of the vortex state for negative magnetic field at  $H_z=-0.05H_{c2}$. The output voltage shows periodic oscillations with several maxima and minima. The global minimum corresponds to the Meissner state (point 1 and inset 1). At a later time two antivortices penetrate the sample, one after the other (insets 2 and 3), leading to local maxima in the $V(t)$ curve just to the left of points 2 and 3. Note that antivortex nucleation does not occur at the outer corner of the sample because the current is minimal there. Rather they nucleate away from this corner, where the current density is maximal (see panel 6 in Fig.\  \ref{fig2} for the current distribution in the sample). These antivortices temporarily reside inside the sample, which leads to local minima in the $V(t)$ curve to the right of points 2 and 3. As time goes on, the antivortices leave the sample, one after the other (insets 4 and 6), producing local maxima in the $V(t)$ curve (points 4 and 6), beyond which the superconducting condensate relaxes towards its initial state. After some time, a new antivortex penetrates the sample and the entire antivortex entry-exit sequence repeats. The time-dependent voltage shown in Fig. 9(b) thus can be regarded as  a periodic sequence of double pulses produced by two antivortices crossing at nearly the same time.  The integrated area of each double pulse is $\int V dt = 2\phi_0$.  Overall, for larger values of the negative field the resistive state is characterized by the periodic entrance of antivortices. 

Figure \ref{fig5} (a) shows the $V(t)$ characteristics of the sample for $H_z=-0.025H_{c2}$, where the critical current of the sample reaches its maximal value (i.e., $j_c=0.425j_0$). Surprisingly, the resistive state in this case is characterized by the motion of both vortices and antivortices, which move in opposite direction without annihilation (see white arrows in panel 1 in Fig.\  \ref{fig5} for their direction of motion). Vortices nucleate at the inner corner of the sample periodically in time and move along the diagonal direction (panels 2-5), whereas antivortices nucleate at the outer edge of the sample and move across it, as we discussed previously in connection with Fig.\  \ref{fig4}. However, antivortices move much faster than vortices, they do not retain their circular shape, and they create a channel with suppressed order parameter. This is due to the more uniform current flow in the regions away from the corner, which was shown to be the condition for the formation of fast-moving kinematic vortices. \cite{Andronov} The speed of these kinematic vortices decreases with further increasing the applied magnetic field.\cite{golib} Coexistence of fast- and slow-moving vortices has also been found in straight superconducting samples. \cite{denisV} Thus, the output voltage is characterized by fast oscillations due to the kinematic vortices, the amplitude of which is periodically modulated due to the slow motion of vortices, as seen from the Fourier power spectrum of the $V(t)$ curve [see Fig.\  \ref{fig5}(b)]        

\begin{figure}[t]
\includegraphics[width=\linewidth]{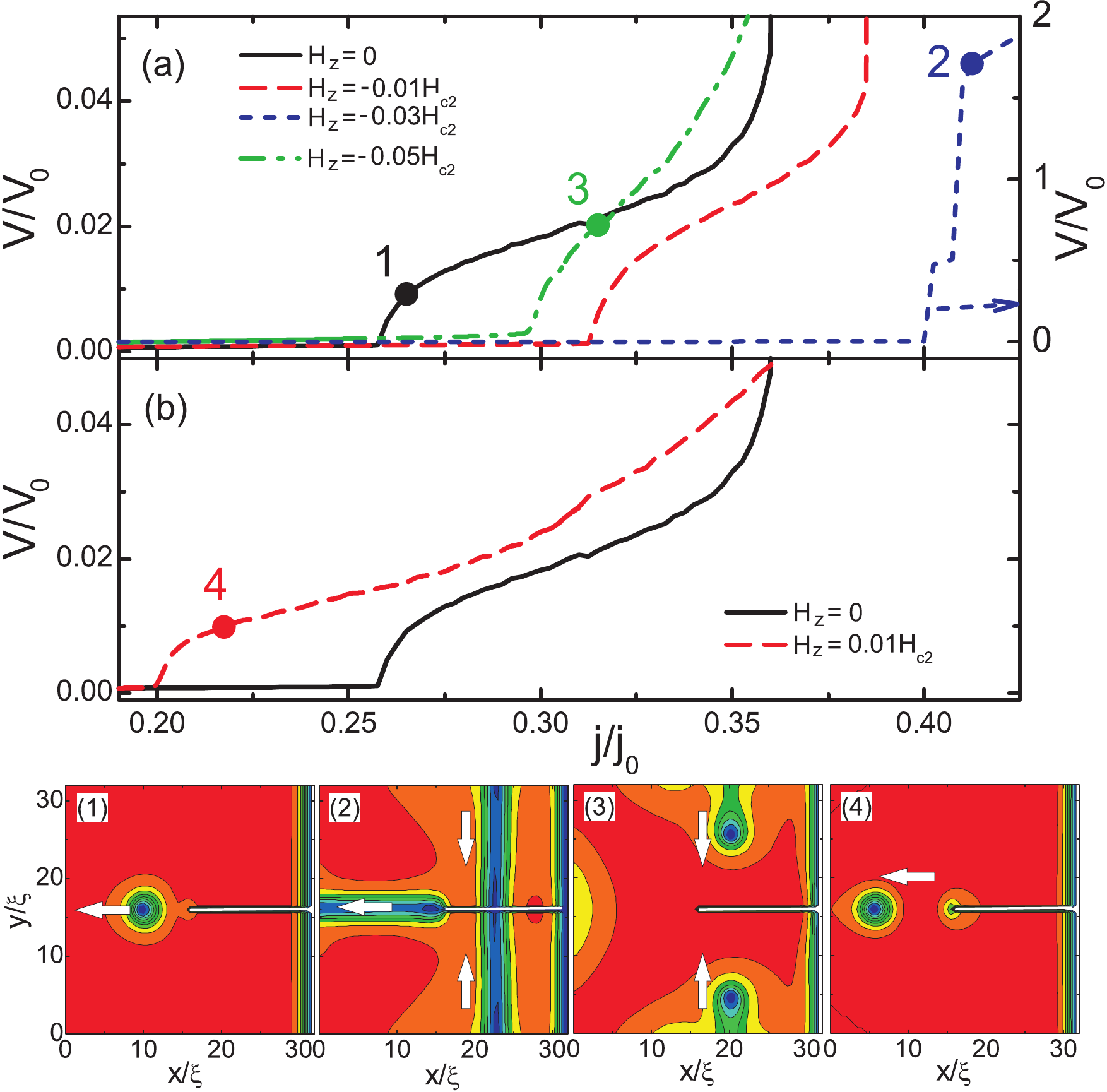}
\caption{\label{fig6}(Color online) $I-V$ characteristics of the sample ($L=32\xi$ and $W=15.5\xi$) with a 180$^{\circ}$ turnaround for different values of (a) negative and (b) positive magnetic field. Panels 1-4 show snapshots of $|\psi|^2$ at field and current values indicated on the $I-V$ curves. White arrows indicate the direction of vortex motion.}
\end{figure}

\begin{figure} [b]
\includegraphics[width=\linewidth]{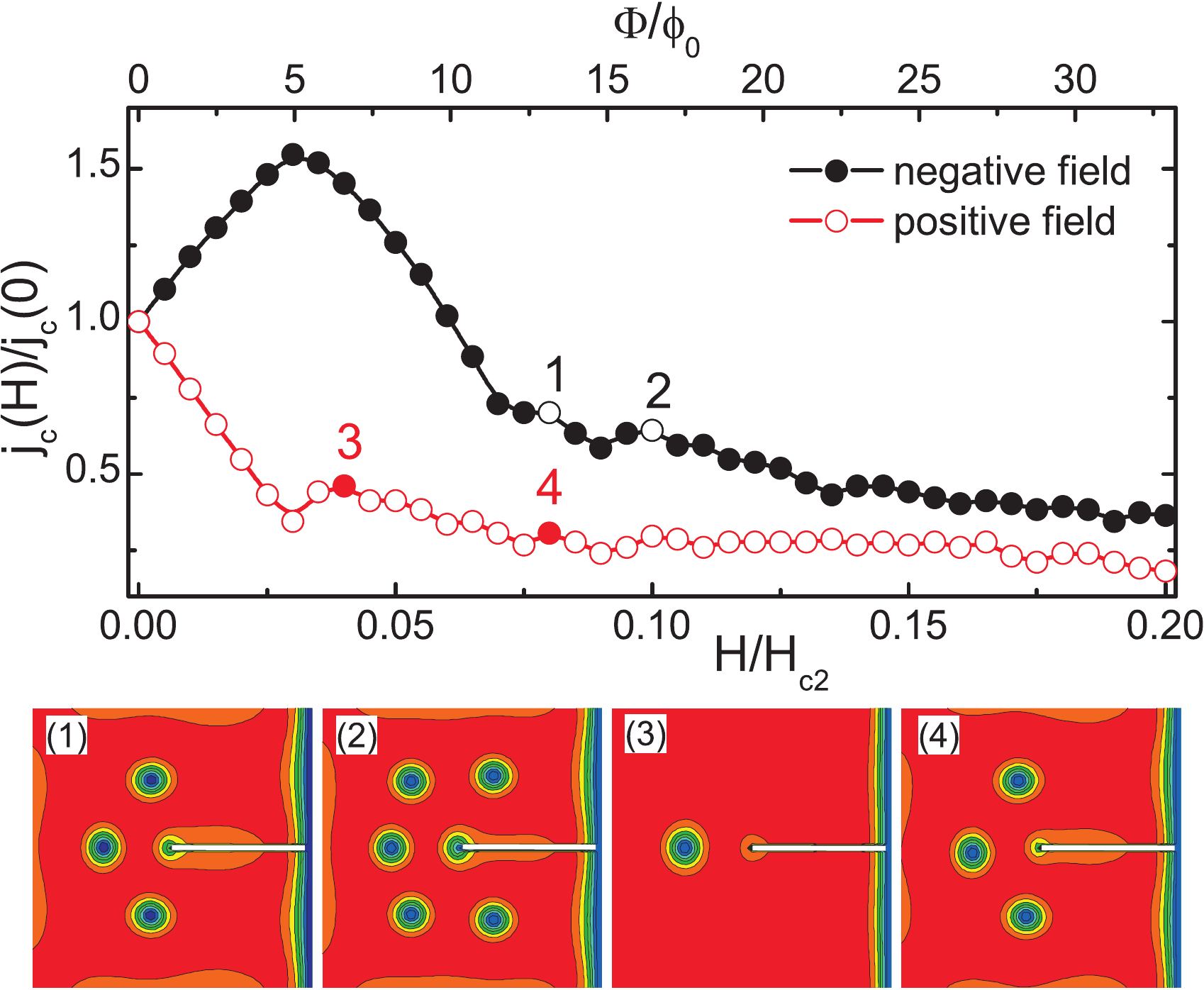}
\caption{\label{fig7}(Color online) The critical current $j_{c}/j_{c}(0)$ of the sample in Fig.\  \ref{fig6} as a function of negative (filled black circles) and positive (open red circles) applied magnetic field. Panels 1-4 show contour plots of $|\psi|^2$ for magnetic field values indicated on the $j_{c}(H)$ curves and for current values just below $j_{c}$.}
\end{figure}

\subsection{Superconducting system with a 180-degree turnaround}

In what follows, we study how the critical current of a superconducting strip with a sharp 180-degree turnaround [see Fig.\  \ref{fig1}(b)] is affected by the external field. As our main results, we plotted in Fig.\  \ref{fig6} the calculated $I-V$ characteristics of the sample with dimensions $W=15.5\xi$ and $L=32\xi$ (i.e., the gap separating two parts of the sample is $\Delta=1\xi$) for different values of negative (a) and positive (b) values of the applied magnetic field. As in the case of the sample with a 90-degree turn, vortices are bound to nucleate at the sharp inner corner in the absence of a magnetic field (see the solid black curve and panel 1 in Fig.\  \ref{fig6}), the motion of which determines the resistive-state transition current. This critical current increases with increasing negative field until some threshold field (see dotted blue curve), above which $j_{c}$ decreases again (dot-dashed green curve). At these values of the field, antivortices penetrate the sample away from the corners (panel 3), leading to energy dissipation in the system. The coexistence of vortices and antivortices is also observed at intermediate values of the field, as shown in panel 2 of Fig.\  \ref{fig6}. For positive direction of the magnetic field, vortices preferentially nucleate at the inner corner (see panel 4) where the screening currents add to the applied current. Thus, the critical current becomes a decreasing function of the field, because vortex entries are shifted to lower currents as compared to the zero-applied-field case. Note that the critical current density $j_{c}$ is always smaller than the one obtained for the 90-degree turn sample due to larger current crowding near the inner corners (compare the $I-V$ curves in Figs. \ref{fig2} and \ref{fig6}).

To determine how the critical current is affected by the applied field and, in particular, to show to what extent the critical current of the sample can be increased by the negative magnetic field, we present in Fig.\  \ref{fig7} the critical current of the system $j_{c}$ as a function of the magnetic field. $j_{c}(H)$ curves show behavior similar to that in the case of a single 90-degree turn (see Fig.\  \ref{fig3}): (i) $j_c$ increases with increasing negative field (filled black circles) until some threshold field, beyond which $j_{c}$ decreases again due to the penetration of antivortices; (ii) for positive direction of the field (open red circles), $j_{c}$ linearly decreases with field because of the increased current density at the corners, which reduces the energy barrier for the nucleation of vortices; (iii) at larger field values different vortex (antivortex) patterns are stabilized in the system (see panels 1-4 in Fig.\  \ref{fig7}) which results in a nonlinear dependence of the critical current on both negative and positive magnetic fields.

\begin{figure}[t]
\includegraphics[width=\linewidth]{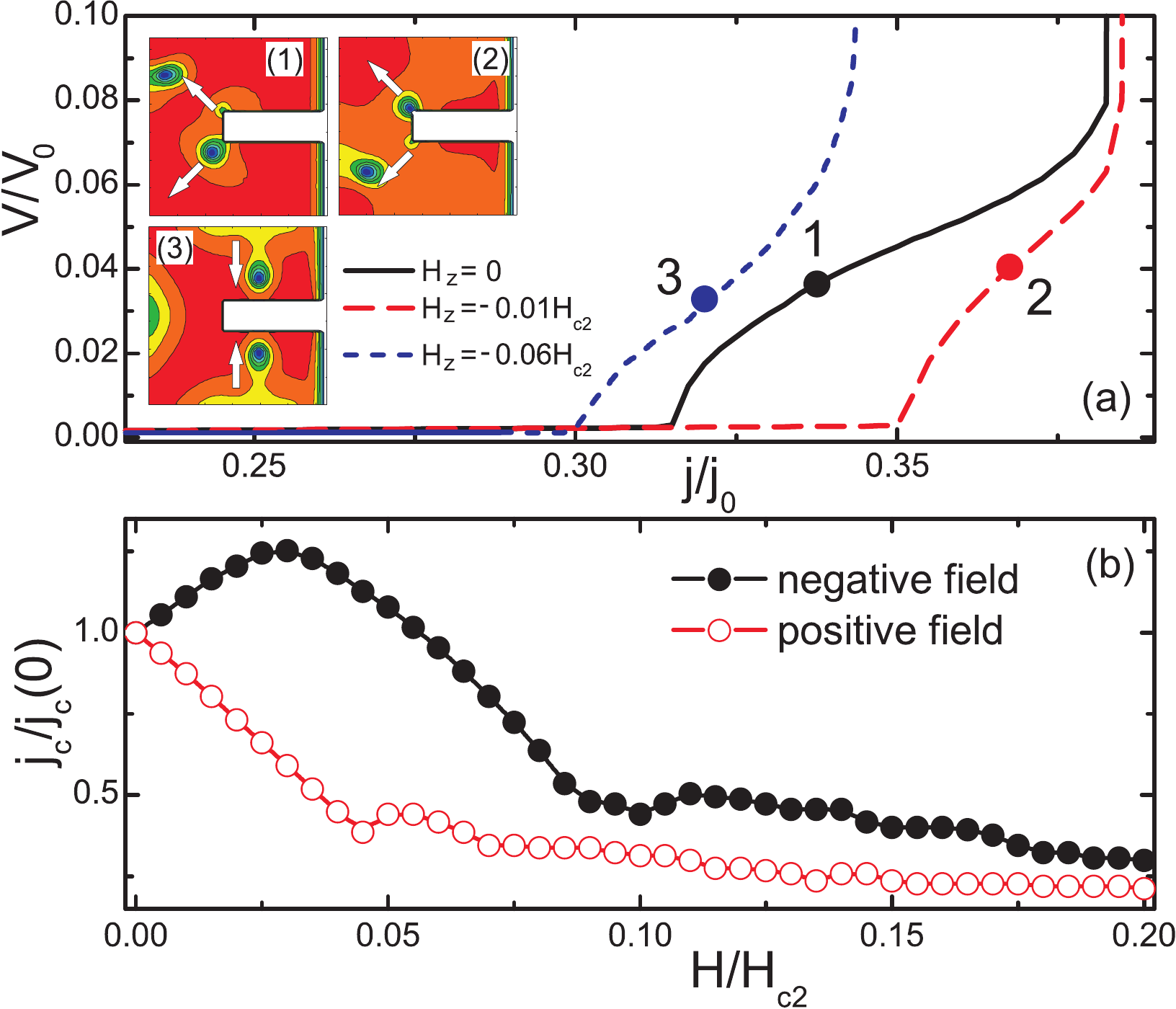}
\caption{\label{fig8}(Color online) (a) $I-V$ characteristics of the sample with $L=32\xi$ and $W=13.5\xi$ (i.e., $\delta=5\xi$) with a 180$^{\circ}$ turnaround for different values of the negative field. Insets show snapshots of $|\psi|^2$ at field and current values indicated on the $I-V$ curves. (b) $j_c(H)$ curves of the sample for negative (solid black circles) and positive (open red circles) magnetic field.}
\end{figure}

Finally, we discuss the effect of the opening $\Delta$ of the sample on our findings [see Fig.\ \ref{fig1}(b)]. Figure \ref{fig8} (a) shows the $I-V$ characteristics of the sample with $L=32\xi$ and $\Delta=5\xi$ for several values of the negative magnetic field.  At zero magnetic field (as well as for positive applied fields) vortices nucleate at the two inner corners of the sample, one after another [see inset 1 in Fig.\  \ref{fig8}(a)], and move along the diagonal direction, as in the case of the 90-degree turn sample. The zero-field critical current of the sample is considerably larger than the one for the sample with a narrow gap (see Fig.\  \ref{fig6}), because of the smaller current crowding near the inner corners. $j_c$ increases with increasing negative magnetic field [filled black circles in Fig.\  \ref{fig8}(b)] until some threshold value, above which $j_c$ is reduced due to the formation of antivortices [inset 3 in Fig.\  \ref{fig8}(a)]. As before, $j_c$ is a decreasing function of the positive magnetic field.   

In conclusion, our numerical simulations confirm the following theoretical findings for superconducting samples with sharp (90- or 180-degree) turns: (i) in the absence of the magnetic field, vortices nucleating at the inner sharp corners determine the resistive-state transition current, which is smaller than the one for a straight strip with no turn due to the current crowding at the inner corners; (ii) the critical current can be increased by applying a perpendicular magnetic field of the right polarity, because of the reduction of the applied current density at the inner corners by the screening currents; (iii) for the other direction of the magnetic field, the critical current decreases linearly with field, because at the nucleation point the field-induced current is in the same direction as the applied current; (iv) for larger values of the magnetic field the critical current decreases (but with slower rate) with the field in either direction; (v) oscillations in the $j_{c}(H)$ curves are observed at larger fields as a consequence of different vortex states, before the system transits to the resistive state.

\section{Summary and Discussion%
\label{Sec_conclusion}}

In this paper we have considered thin (thickness $d < \lambda$) and narrow (width $W\ll \Lambda = 2\lambda^2/d$ and $\xi \ll W$) superconducting strips with  sharp turns  in the middle.  Using a London-model description of vortices, we showed theoretically that when an applied current $I$ flows through the strips shown in Figs.\ \ref{90Fig}-\ref{sharp180Fig} and makes {\it left} turns, the critical sheet current, dominated by the onset of {\it vortex} nucleation at the sharp inner corners, is suppressed by current crowding.  We have shown that this critical current further decreases linearly as a positive magnetic field $H_z$ is applied.  However, the critical current {\it increases} linearly as the applied field becomes more negative.  The maximum critical current [see Eq.\ (\ref{Kcp})] is reached at the field [see Eq.\ (\ref{Hp})] when the critical current for vortex nucleation at a sharp inner corner becomes equal to the critical current for {\it antivortex} nucleation along one of the straight sides far from the outer corners.  
The behavior when the current makes {\it right} turns can be understood using simple symmetry arguments, as shown in  Fig.\ \ref{Kcv&KcaFig90}(b).

To confirm these theoretical predictions, we performed  numerical simulations within the time-dependent Ginzburg-Landau (TDGL) theory. Figures \ref{fig3}, \ref{fig7}, and \ref{fig8} show results for the field dependence of the critical current in general agreement with the generic behavior predicted in Fig.\  \ref{Kcv&KcaFig90}.  The predicted maximum enhancements of the critical current in negative applied fields calculated from Eq.\ (\ref{Kcp}) using the expressions for $R$ and $\sigma$ given in Appendix \ref{details} are $K_{cp}/K_{cv}(0) = 1.34$, 1.53, and 1.35, which are in reasonable agreement with the peak values of $j_c/j_c(0)$, in Figs.\ \ref{fig3}, \ref{fig7}, and \ref{fig8} respectively.  On the other hand, the fields at which the maximum enhancements are predicted to occur, calculated from Eq.\ (\ref{Hp}),  $H_{max}/H_{c2} = -0.012,\;-0.016, {\rm and} -0.014$, are not in agreement with the positions of the peaks in Figs.\ \ref{fig3}, \ref{fig7}, and \ref{fig8}, $H_{max}/H_{c2} = -0.025,\;-0.03, {\rm and} -0.03$, respectively.  
This discrepancy evidently arises from the quantitative inaccuracy of the London-model description and its handling of the vortex core  on the length scale of $\xi$.

Another difference between the predictions of Secs.\ I-III and the TDGL simulations is that the peak of the critical current vs field calculated using the London-model approach and shown in Fig.\ \ref{Kcv&KcaFig90}(a) resembles an inverted ``V", whereas all the simulations show that this peak is rounded.  The explanation of this difference is that the calculations of Secs.\ I-III were performed in the limit of very large $W$ relative to  $\xi$, whereas the simulations of Sec.\ \ref{TDGL} were carried out for $W/\xi$ equal to either 15.5 (Figs.\ \ref{fig3} and \ref{fig7}) or 13.5 (Fig.\ \ref{fig8}).  Simulations for 90-degree turns with different values of $W/\xi$ (not shown here) have revealed that the peak in the critical current vs field becomes sharper as the ratio $W/\xi$ increases.  

The above-described effects apply to all asymmetric geometries of thin and narrow superconducting films.  We thus predict that, upon application of a suitable perpendicular magnetic field, the critical current in such patterns will depend strongly upon the direction of the current, allowing rectification of ac currents and producing a diode effect.  Another possible application of the field-induced enhancement of the critical current is in superconducting nanowire single-photon detectors.\cite{Goltsman01,Sobolewski03,Yang09}  
For the case of detectors with a two-dimensional meander layout  (a ``boustrophedonic'' pattern), which have both left- and right-handed 180-degree turns, our results lead to the prediction that application of a perpendicular magnetic field would reduce the critical current.  
On the other hand, we predict that application of a negative applied field would increase the critical current in detectors using thin and narrow superconducting strips in a spiral layout\cite{Jaycox81} where the current makes only left turns.  Fabrication of such spiral layouts, however, would be more difficult because of the need to electrically connect the center of the superconducting spiral through a hole in an insulating layer to a strip in a second superconducting layer.

\section*{ACKNOWLEDGMENTS} 
This research, supported in part by the U.S.\ Department of
Energy, Office of Basic Energy Science, Division of Materials
Sciences and Engineering, was performed in part at
the Ames Laboratory, which is operated for the U.S.\ Department
of Energy by Iowa State University under Contract No.
DE-AC02-07CH11358.  
This work also was supported in part by the Flemish Science Foundation (FWO-Vlaanderen) and the Belgian Science Policy (IAP). G.R.B. acknowledges individual support from FWO-Vlaanderen.

\appendix

\section{Details\label{details}}

\subsection{Conformal mappings}

To calculate the functions needed for the calculation of the stream functions, sheet currents, and Gibbs free energies, we follow the approach given in Ref.\ \onlinecite{Clem11} by first finding conformal mappings that map the upper half $w$-plane ($w = u +iv$) into the regions in the $\zeta$-plane ($\zeta = x+iy$) corresponding to the films shown in Figs.\ \ref{90Fig}-\ref{sharp180Fig}.   

\subsubsection{$\zeta(w)$ for the sharp 90-degree turn}

The following mapping describes the sharp 90-degree turn shown in Fig.\ \ref{90Fig}:
\begin{eqnarray}
\frac{d\zeta(w)}{dw}  &=& \frac{iW\sqrt{w-1}}{\pi w\sqrt{w+1}},\label{dzetadws90}\\
\zeta(w)&=&\frac{W}{\pi}[\pi +i \cosh^{-1}(w)+\cosh^{-1}(-\frac{1}{w})],\label{zetaofws90}
\end{eqnarray}
and $w(\zeta)$, the inverse of $\zeta(w)$, must be obtained numerically.  [This choice of $w$ differs by 1 from that used in Eqs.\ (103) and (104) in Ref.\ \onlinecite{Clem11}.]
Referring to Fig.\ \ref{90Fig}, the letters A-F indicate mappings as follows: (A) $x = 0$, $y = +\infty$, $w= -\infty$, (B) $x=y = 0$, $w = -1$, (C) $x=+\infty$, $y = 0$, $w = -\epsilon$, (D) $x = +\infty$, $y = 
W$, $w = +\epsilon$, (E) $x = y=W$, $w= 1$, and (F) $x = W$, $y = +\infty$, $w = +\infty$, where $\epsilon$ is a positive infinitessimal. 

Doing an expansion about the point E, we find that for $\zeta$  a short distance $\delta \ll W$ away from E along the diagonal between E and B [$\zeta = W(1+i)-\delta e^{i\pi/4}$] the corresponding value of $w$ is to lowest order
\begin{equation}
w(\zeta) = 1+i\Big(\frac{3\pi\delta}{W\sqrt{2}}\Big)^{2/3}.
\label{wexps90}
\end{equation}

\subsubsection{$\zeta(w)$ for the sharp  rectangular 180-degree turnaround}

The following mapping describes the  sharp rectangular 180-degree turn shown in Fig.\ \ref{rect180Fig}:
\begin{eqnarray}
\frac{d\zeta(w)}{dw}  &=& \frac{iA\beta\sqrt{w^2-\alpha^2}}{(w^2-1)\sqrt{w^2-\beta^2}},\label{dzetadwsr180}\\
\zeta(w)&=&iA[(1-\alpha^2)\Pi(\sin^{-1}\frac{w}{\alpha},\alpha^2,\frac{\alpha}{\beta})\nonumber \\ &&-F(\sin^{-1}\frac{w}{\alpha},\frac{\alpha}{\beta})],\label{zetaofwsr180}
\end{eqnarray}
where 
$F(\phi,k)$ and $\Pi(\phi,n,k)$ are elliptic integrals of the first and third kind\cite{Gradshteyn00} of argument $\phi$, modulus $k$, and parameter $n$, 
and $w(\zeta)$, the inverse of $\zeta(w)$, must be obtained numerically.  
Referring to Fig.\ \ref{rect180Fig}, the letters A-J indicate mappings as follows: (A) $x = W$, $y = 0$, $w= -\infty$, (B) $x=W$, $y = W+h$, $w = -\beta$, (C) $x=-\infty$, $y = W+h$, $w = -1-\epsilon$, (D) $x = -\infty$, $y = h$, $w = -1+\epsilon$, (E) $x =0$, $ y=h$, $w= -\alpha$, (O) $x = y = 0$, $w = 0$, (F) $x = 0$, $y = -h$, $w = \alpha$, (G) $x = -\infty,$ $y = -h$, $w = 1-\epsilon,$ (H) $x = -\infty$, $y = -W-h$, $w = 1+\epsilon,$ (I) $x = W$, $y = -W-h$, $w = \beta$, and (J) $x = W$, $y = 0$, $ w = +\infty$, where $\epsilon$ is a positive infinitessimal.  This mapping yields three equations that determine the values of $A$, $\alpha$, and $\beta$ for given values of $W$ and $h$: 
\begin{eqnarray}
W&=& \frac{\pi A\beta\sqrt{1-\alpha^2}}{2\sqrt{\beta^2-1}},\label{WEq}\\
W&=&\frac{A}{(\beta^2-1)}[(\beta^2-\alpha^2){\bm K}(\frac{\sqrt{\beta^2-\alpha^2}}{\beta})\nonumber\\
&&-(1-\alpha^2){\bm \Pi}(\frac{\beta^2-1}{\beta^2},\frac{\sqrt{\beta^2-\alpha^2}}{\beta})],\label{WpEq}\\ 
h&=&A[{\bm K}(\frac{\alpha}{\beta})-(1-\alpha^2){\bm \Pi}(\alpha^2,\frac{\alpha}{\beta})],\label{hEq}
\end{eqnarray}
where 
$\bm K(k)$ and $\bm \Pi(n,k)$ are complete elliptic integrals of the first and third kind\cite{Gradshteyn00} of modulus $k$ and parameter $n$.  For example, solving  Eqs.\ (\ref{WEq})-(\ref{hEq}) for $h/W = 0.15$, as shown in Fig.\ \ref{rect180Fig}, yields $A = 0.609 W$, $\alpha = 0.534$, and $\beta = 1.703$.  

Doing an expansion about the point F for $\zeta$ along the diagonal between F and I, we find that for $\zeta$ a short distance $\delta \ll W$ away from F ($\zeta = -ih+\delta e^{-i\pi/4}$) the corresponding value of $w$ is to lowest order
\begin{equation}
w= \alpha+i\frac{(1-\alpha^2)(\beta^2-\alpha^2)^{1/3}}{2\alpha^{1/3}(\beta^2-1)^{1/3}}\Big(\frac{3\pi\delta}{2W}\Big)^{2/3}.
\label{wexpsr180}
\end{equation}

\subsubsection{$\zeta(w)$ for the sharp  180-degree turnaround}

The following mapping describes the  sharp 180-degree turnaround shown in Fig.\ \ref{sharp180Fig}:
\begin{eqnarray}
\frac{d\zeta(w)}{dw}  &=& i\frac{2W\sqrt{\beta^2-1}}{\pi} \frac{w}{(w^2-1)\sqrt{w^2-\beta^2}},\label{dzetadws180}\\
\zeta(w)&=&W-i\frac{2}{\pi}W\tan^{-1}\Big(\frac{\sqrt{\beta^2-1}}{\sqrt{w^2-\beta^2}}\Big),\label{zetaofws180}\\
w(\zeta)&=&\!\!\!\!\Big[\!\coth^2\!\!\Big(\!\frac{\pi(\zeta\!-\!W)}{2W}\!\Big)\!-\!\beta^2{\rm csch}^2\!\Big(\!\frac{\pi(\zeta\!-\!W)}{2W}\!\Big)\!\Big]^{1/2},\label{wofzetas180}
\end{eqnarray}
where 
$\beta = \cosh(\pi/2)$, and the root in Eq. (\ref{wofzetas180}) must be chosen such that ${\rm Im}w(\zeta)\ge 0$.   
Referring to Fig.\ \ref{sharp180Fig}, the letters A-J indicate mappings as follows: (A) $x = W$, $y = 0$, $w= -\infty$, (B) $x=W$, $y = W$, $w = -\beta$, (C) $x=-\infty$, $y = W$, $w = -1-\epsilon$, (D) $x = -\infty$, $y = \epsilon$, $w = -1+\epsilon$, (O) $x = y = 0$, $w = 0$, (G) $x = -\infty,$ $y = -\epsilon$, $w = 1-\epsilon,$ (H) $x = -\infty$, $y = -W$, $w = 1+\epsilon,$ (I) $x = W$, $y = -W$, $w = \beta$, and (J) $x = W$, $y = 0$, $ w = +\infty$, where $\epsilon$ is a positive infinitessimal.  

Doing an expansion about the origin O for $\zeta$ along the $x$ axis, we find that for $\zeta$ a short distance $\delta \ll W$ away from O ($\zeta = \delta$) the corresponding value of $w$ is to lowest order
\begin{equation}
w= +i\Big(\frac{\pi\delta}{W}\coth\frac{\pi}{2}\Big)^{1/2}.
\label{wexps180}
\end{equation}

\subsection{Stream function for the applied current\label{SI}}

To obtain the stream function for the applied current, we start with a complex potential ${\cal G}_w(w)$ describing the current flow in the $w$-plane.  The corresponding complex potential  ${\cal  G}_\zeta(\zeta)={\cal G}_w(w(\zeta))$  describes the current flow in the $\zeta$-plane. For all values of $x$ and $y$ in the strip,  $\bm K_I(x,y) = \nabla \times \bm S_I(x,y)$, where $\bm S_I(x,y) = \hat z S_{Iz}(x,y)$ is the stream function, \cite{Clem11} and $S_{Iz}(x,y) = \Im{\cal G}_\zeta(x+iy)$, the imaginary part of ${\cal G}_\zeta(\zeta)$. In general, $K_x(x,y) = \partial S(x,y)/\partial y$ and $K_y(x,y) = -\partial S(x,y)/\partial x$.

\subsubsection{$S_{Iz}(x,y)$ for the sharp 90-degree turn}

For the sharp 90-degree turn, we start with the complex potential 
\begin{equation}
{\cal G}_w(w)=-\frac{I}{\pi}\ln w,
\label{calGws90}
\end{equation}
which describes the flow of current $I$ in the upper half $w$-plane from a source at $w = \infty$ to a drain at $w = 0$.  In the $\zeta$-plane the same complex potential is
\begin{equation}
{\cal G}_\zeta(\zeta)={\cal G}_w(w(\zeta))=-\frac{I}{\pi}\ln[w(\zeta)],
\label{calGzetas90}
\end{equation}
where $w(\zeta)$ is the inverse of $\zeta(w)$ given in Eq.\ (\ref{zetaofws90}). The corresponding stream function $S_{Iz}(x,y)$ is shown in Fig.\ \ref{90Fig}(a).

\subsubsection{$S_{Iz}(x,y)$ for the sharp rectangular 180-degree turnaround}

For the sharp rectangular 180-degree turnaround, the  complex potential 
\begin{equation}
{\cal G}_w(w)=\frac{I}{\pi}\ln\Big(\frac{w-1}{w+1}\Big)
\label{calGwsr180}
\end{equation}
describes the flow of current $I$ in the upper half $w$-plane from a source at $w = 1$ to a drain at $w = -1$.  In the $\zeta$-plane the same complex potential is
\begin{equation}
{\cal G}_\zeta(\zeta)={\cal G}_w(w(\zeta))=\frac{I}{\pi}\ln\Big(\frac{w(\zeta)-1}{w(\zeta)+1}\Big),
\label{calGzetasr180}
\end{equation}
where $w(\zeta)$ is the inverse of $\zeta(w)$ given by Eq.\ (\ref{zetaofwsr180}). The corresponding stream function $S_{Iz}(x,y)$ is shown in Fig.\ \ref{rect180Fig}(a).

\subsubsection{$S_{Iz}(x,y)$ for the sharp 180-degree turnaround}

For the sharp 180-degree turnaround, the  complex potential is again given by Eq.\ (\ref{calGwsr180}) or (\ref{calGzetasr180}), but with  $w(\zeta)$ given in Eq.\ (\ref{zetaofwsr180}). The corresponding stream function $S_{Iz}(x,y)$ is shown in Fig.\ \ref{sharp180Fig}(a).

\subsection{Stream function for the field-induced current\label{SH}}

 Because   $\nabla \cdot \bm K_H = 0$, the sheet current induced  by an applied field $\bm H = \hat z H_z$ can be expressed as $\bm K_H = \nabla \times \bm S_H$, where $\bm S_H = \hat z S_{Hz}$ is the stream function.  (This corresponds to the Meissner response of the sample.) 
This contribution to the total sheet-current density $\bm K$ must obey the London equation $\bm 
K_H = -(2/\mu_0 \Lambda)[\bm A_H +(\phi_0/2\pi)\nabla \gamma_H],$ where $\bm A_H$ is the vector potential associated with the applied field ($\bm B = \mu_0 \bm H =  \nabla \times \bm A_H$), and $\gamma_H$ is the corresponding phase of the order parameter.  There are many possible choices for $\bm A_H$, but whatever choice is made, an expression for $\gamma_H$ must be found that makes the gauge-invariant combination $\bm A_H +(\phi_0/2\pi)\nabla \gamma_H$ obey the same boundary conditions as does $\bm K_H$, namely that (i) $\nabla \cdot \bm K_H = 0$, (ii) $\bm K_H \cdot \hat n = 0$, where $\hat n$ is an outward normal to the sample, and (iii) $\bm K_H$ carries no net current. 

We begin by separating $\bm K_H$ into two contributions, $\bm K_H = \bm K_A + \bm K_\gamma$, which obey $\nabla \times \bm K_A =   -(2/\mu_0 \Lambda)\bm B$ and $\nabla \times \bm K_\gamma = 0$, and we also separate the stream function into two contributions, $\bm S_H=\bm S_A+\bm S_\gamma$.  With the choice $\bm A = \hat y B_z(x-W/2)$, we obtain \begin{equation}
\bm K_A = -\hat y (2H_z/\Lambda)(x-W/2)
\label{KA}
\end{equation}
 and 
\begin{equation}
\bm S_{A}=\hat z S_{Az}= \hat z (H_z/\Lambda)x(x-W).
\label{SA}
\end{equation}
Although $\bm K_A$ has the properties required of $\bm K_H$  that  (i) $\nabla \cdot \bm K_A$ = 0, (ii) $\bm K_A \cdot \hat n = 0$ along some of the boundaries shown in Figs.\ \ref{90Fig}-\ref{sharp180Fig}, and (iii) the integral of $\bm K_A$ across the strip is zero (the strip carries no net current), it does not satisfy the requirements of zero normal component along all the boundaries of the sample. We therefore need to find $\bm K_\gamma$ such that $\bm K_H = \bm K_A + \bm K_\gamma$ satisfies all the boundary conditions.

We may represent the sheet-current density $\bm K_\gamma = \hat x K_{\gamma x}(x,y) + \hat y K_{\gamma y}(x,y)$ by the analytic function ${\cal K}_\gamma(\zeta) = K_{\gamma x}(x,y) -i K_{\gamma y}(x,y)$  of the complex variable $\zeta = x+iy$. The equations  $\nabla \times \bm K_\gamma = 0$ and $\nabla \cdot \bm K_\gamma = 0$ are equivalent to the Cauchy-Riemann equations obeyed by analytic functions.  The corresponding complex potential is $\cal G_\gamma(\zeta)$, where ${\cal K}_\gamma(\zeta) = d{\cal G}_\gamma(\zeta)/d\zeta$, and the stream function is $\bm S_\gamma = \hat z S_{\gamma z}$, where $S_{\gamma z}(x,y) = {\rm Im}{\cal G}_\gamma(x+iy)$.  For each of the geometries shown in Figs.\ \ref{90Fig}-\ref{sharp180Fig}, the problem then reduces to finding the appropriate $S_{\gamma z}(x,y)$ that corrects for the failure of $S_{Az}$ to satisfy all the boundary conditions. 

\subsubsection{$S_{Hz}(x,y)$ for the sharp 90-degree turn}

For the sharp 90-degree turn, the stream function, $S_{Hz} = S_{Az}+S_{\gamma z}$, is given by 
\begin{eqnarray}
S_{Hz}(x,y)&=&\frac{H_z}{\Lambda}\Big\{x(x-W)\nonumber\\
&+&\frac{1}{\pi}{\rm Im}\int_{-1}^1\frac{x(u)[x(u)-W]}{w(x+iy)-u}du\Big\},
\label{SHzs90}
\end{eqnarray}
where $x(u) = {\rm Re}\zeta(u+i\epsilon)$, obtained from Eq.\ (\ref{zetaofws90})], is
\begin{equation}
x(u) = (W/\pi)[\cos^{-1}(-u)+\cosh^{-1}(1/|u|)], 
\label{xofu}
\end{equation}
and $w(x+iy)$ is obtained from the  inverse of Eq.\ (\ref{zetaofws90}).
The resulting stream function is shown in Fig.\ \ref{90Fig}(b) as a contour plot.  

\subsubsection{$S_{Hz}(x,y)$ for the sharp rectangular 180-degree turnaround}

For the sharp rectangular 180-degree turnaround, the stream function, $S_{Hz} = S_{Az}+S_{\gamma z}$, is given by 
\begin{eqnarray}
S_{Hz}(x,y)&=&\frac{H_z}{\Lambda}\Big\{x(x-W)\nonumber\\
+&&\!\!\!\!\!\!\!\!\!\!\!\!\frac{2}{\pi}{\rm Im}\!\!\int_{\alpha}^\beta\!\!\frac{x(u)[x(u)\!-\!W]w(x\!+\!iy)}{w^2(x\!+\!iy)-u^2}du\Big\},
\label{SHzsr180}
\end{eqnarray}
where $x(u) = {\rm Re}\zeta(u+i\epsilon)$ is obtained from Eq.\ (\ref{zetaofwsr180}) and $w(x+iy)$ is obtained from the inverse of Eq.\ (\ref{zetaofwsr180}).
The resulting stream function is shown in Fig.\ \ref{rect180Fig}(b) as a contour plot.  

\subsubsection{$S_{Hz}(x,y)$ for the sharp 180-degree turnaround}

For the sharp 180-degree turnaround, the stream function, $S_{Hz} = S_{Az}+S_{\gamma z}$, is given by 
\begin{eqnarray}
S_{Hz}(x,y)&=&\frac{H_z}{\Lambda}\Big\{x(x-W)\nonumber\\
+\!&&\!\!\!\!\!\!\!\!\!\!\!\frac{2}{\pi}{\rm Im}\!\!\int_{0}^\beta\!\!\!\frac{x(u)[x(u)\!-\!W]w(x\!+\!iy)}{w^2(x\!+\!iy)-u^2}du\Big\},
\label{SHzs180}
\end{eqnarray}
where $x(u) = {\rm Re}\zeta(u+i\epsilon)$ is obtained from Eq.\ (\ref{zetaofws180})] and $w(x+iy)$ is obtained from  Eq.\ (\ref{wofzetas180}).  
The resulting stream function is shown in Fig.\ \ref{sharp180Fig}(b) as a contour plot.  

\subsection{Applied sheet-current density near the sharp inner corners}

As suggested by the current crowding  shown in Figs.\ \ref{90Fig}(a)-\ref{sharp180Fig}(a), the sheet-current density generated by the applied current diverges upon approaching the sharp inner corners.  In the following we calculate $K_{I\delta}$, the counterclockwise component of the sheet-current density at a distance $\delta$ ($\delta \ll W$) from the corner.  We use the complex potentials given in Sec.\  \ref{SI} to evaluate  ${\cal K}_\zeta = d{\cal G}/d\zeta = K_x - i K_y$.  

\subsubsection{$K_{I\delta}$ for the sharp 90-degree turn}

Starting from  Eqs.\ (\ref{calGzetas90}) and (\ref{dzetadws90}), and doing an expansion about the point E using Eq.\ (\ref{wexps90}), we find for $\zeta$  a short distance $\delta \ll W$ away from E along the diagonal between E and B [$\zeta = W(1+i)-\delta e^{i\pi/4}$] that the counterclockwise component of the sheet current density is to lowest order
\begin{equation}
K_{I\delta} = K_I\Big(\frac{4W}{3\pi\delta}\Big)^{1/3}.
\label{KIdeltas90}
\end{equation} 
Integrating the Lorentz force $\phi_0 K_{I\delta}$, we find that the work done by the source of the current $K_I$ in moving the vortex from the corner to the position $\delta$ is 
\begin{equation}
W_I(\delta) = \phi_0 K_I \Big(\frac{W}{2\pi}\Big)^{1/3}(3\delta)^{2/3}.
\label{WIs90}
\end{equation}

\subsubsection{$K_{I\delta}$ for the sharp rectangular 180-degree turnaround}

Starting from  Eqs.\ (\ref{calGzetasr180}) and (\ref{dzetadwsr180}), and 
doing an expansion about the point F using Eq.\ (\ref{wexpsr180}), we find for $\zeta$  a short distance $\delta \ll W$ from F along the diagonal between F and I ($\zeta = -ih+\delta e^{-i\pi/4}$) that the counterclockwise component of the sheet current density is to lowest order
\begin{equation}
K_{I\delta} = K_I\frac{(\beta^2-\alpha^2)^{1/3}}{(\beta^2-1)^{1/3}}\Big(\frac{2W}{3\pi\alpha\delta}\Big)^{1/3}.
\label{KIdeltasr180}
\end{equation} 
Integrating the Lorentz force $\phi_0 K_{I\delta}$, we find that the work done by the source of the current $K_I$ in moving the vortex from the corner to the position $\delta$ is 
\begin{equation}
W_I(\delta) = \phi_0 K_I \frac{(\beta^2-\alpha^2)^{1/3}}{(\beta^2-1)^{1/3}}\Big(\frac{W}{\pi\alpha}\Big)^{1/3}\Big(\frac{3\delta}{2}\Big)^{2/3}.
\label{WIsr180}
\end{equation}

\subsubsection{$K_{I\delta}$ for the sharp  180-degree turnaround}

Starting from  Eqs.\ (\ref{calGzetasr180}) and (\ref{dzetadws180}), and 
doing an expansion about the origin O using Eq.\ (\ref{wexps180}), we find for $\zeta$  a short distance $\delta \ll W$ from O along the $x$ axis ($\zeta = \delta $) that the counterclockwise component of the sheet current density is to lowest order
\begin{equation}
K_{I\delta} = K_I\Big(\frac{W}{\pi\delta}\coth\frac{\pi}{2}\Big)^{1/2}.
\label{KIdeltas180}
\end{equation}
Integrating the Lorentz force $\phi_0 K_{I\delta}$, we find that the work done by the source of the current $K_I$ in moving the vortex from the corner to the position $\delta$ is 
\begin{equation}
W_I(\delta) = 2\phi_0 K_I \Big(\frac{W\delta}{\pi}\coth\frac{\pi}{2}\Big)^{1/2}.
\label{WIs180}
\end{equation}

\subsection{Field-induced sheet-current density near the sharp inner corners}

As suggested by the current crowding  shown in Figs.\ \ref{90Fig}(b)-\ref{sharp180Fig}(b), the sheet-current density induced by the applied magnetic field (the Meissner current) diverges upon approaching the sharp inner corners.  In the following we calculate $K_{H\delta}$, the counterclockwise component of the sheet-current density at a distance $\delta$ ($\delta \ll W$) from the corner.  We obtain $K_{H\delta}$ by evaluating  $\bm K_H = \nabla \times \bm S_H$, where $\bm S_H = \hat z S_{Hz}$ is one of the stream functions given in Sec.\  \ref{SH}.  For each of the three cases, the divergence of $K_{H\delta}$ arises from the integrals given in Eqs.\ (\ref{SHzs90}),   (\ref{SHzsr180}), and (\ref{SHzs180}).  

\subsubsection{$K_{H\delta}$ for the sharp 90-degree turn}

Starting from  Eqs.\ (\ref{SHzs90}) and (\ref{dzetadws90}), and doing an expansion about the point E using Eq.\ (\ref{wexps90}), we find for $\zeta$  a short distance $\delta \ll W$ away from E along the diagonal between E and B [$\zeta = W(1+i)-\delta e^{i\pi/4}$] that the counterclockwise component of the field-induced sheet-current density is to lowest order
\begin{equation}
K_{H\delta} = \sigma \frac{H_zW}{\Lambda}\Big(\frac{4W}{3\pi\delta}\Big)^{1/3},
\label{KHdeltas90}
\end{equation}
where, with $x(u)$ given by Eq. (\ref{xofu}),
\begin{equation} 
\sigma = \frac{1}{W^2}\int_{-1}^1\frac{x(u)[x(u)-W]}{(1-u)^2}du = \frac{8\bm G}{\pi^2}=0.742
\label{sigmas90}
\end{equation}
and $\bm G = 0.915965...$ is Catalan's constant.
Integrating the Lorentz force $\phi_0 K_{H\delta}$, we find that the work done by the source of the applied field $H_z$ in moving the vortex from the corner to the position $\delta$ is 
\begin{equation}
W_H(\delta) =  \phi_0 \sigma\frac{H_zW}{\Lambda} \Big(\frac{W}{2\pi}\Big)^{1/3}(3\delta)^{2/3}.
\label{WHs90}
\end{equation}

\subsubsection{$K_{H\delta}$ for the sharp rectangular 180-degree turnaround}

\begin{figure}
\includegraphics[width=8cm]{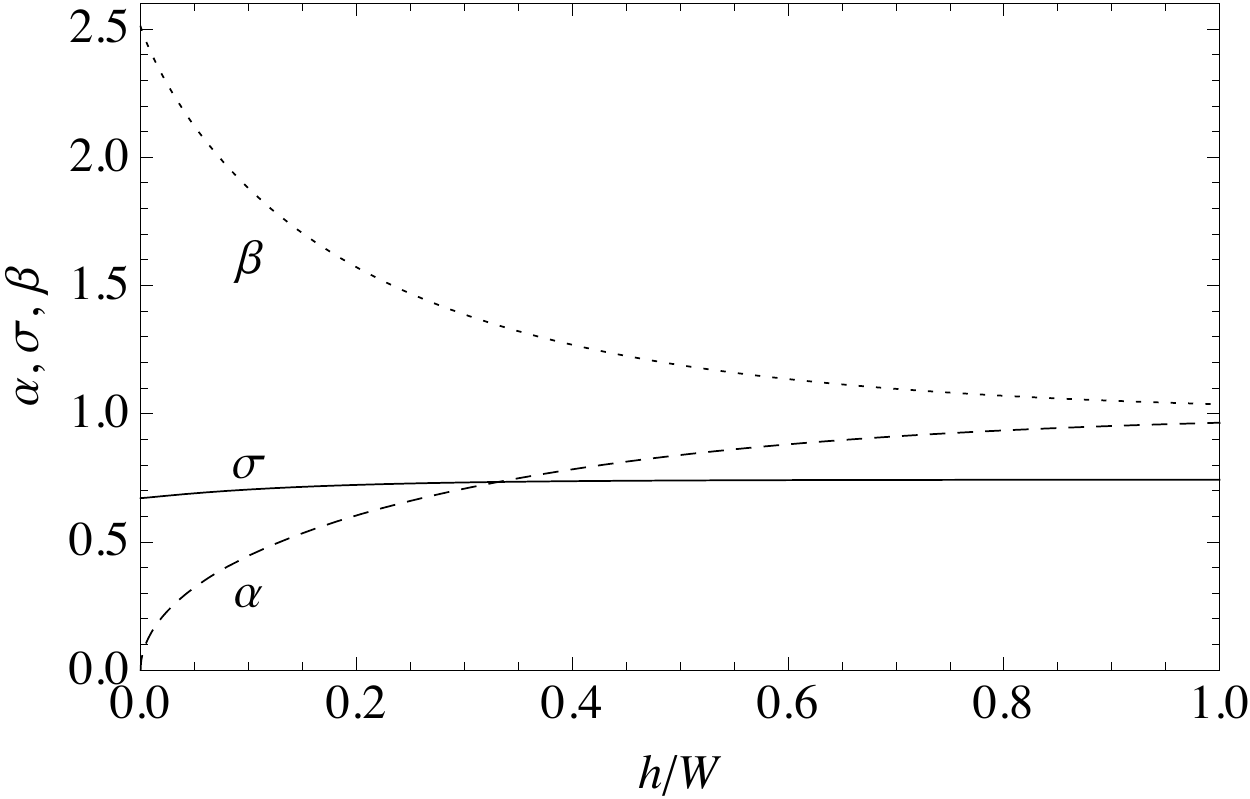}
\caption{%
Parameters $\alpha$, $\beta$, and the slope parameter $\sigma$ [Eq.\ (\ref{sigmasr180})] as functions of $h/W$ for the sharp rectangular 180-degree turnaround shown in Fig.\ \ref{rect180Fig}.  The strip width is $W$, and the gap opening is $\Delta = 2h$.   }
\label{sigmaPlot}
\end{figure}

Starting from  Eqs.\ (\ref{SHzsr180}) and (\ref{dzetadwsr180}), and 
doing an expansion about the point F using Eq.\ (\ref{wexpsr180}), we find for $\zeta$  a short distance $\delta$ ($h \ll \delta \ll W$) from F along the diagonal between F and I ($\zeta = -ih+\delta e^{-i\pi/4}$) that the counterclockwise component of the field-induced  sheet-current density is to lowest order
\begin{equation}
K_{H\delta} =\sigma \frac{H_zW}{\Lambda}\frac{(\beta^2-\alpha^2)^{1/3}}{(\beta^2-1)^{1/3}}\Big(\frac{2W}{3\pi\alpha\delta}\Big)^{1/3},
\label{KHdeltasr180}
\end{equation} 
where, with $x(u)$ obtained from Eq.\ (\ref{zetaofwsr180}),
\begin{equation} 
\sigma = \frac{(1-\alpha^2)}{W^2}\int_{\alpha}^\beta\frac{(\alpha^2+u^2)x(u)[x(u)-W]}{(\alpha^2-u^2)^2}du .
\label{sigmasr180}
\end{equation}
 For example, solving  Eqs.\ (\ref{WEq})-(\ref{hEq}) for $h/W = 0.15$, as shown in Fig.\ \ref{rect180Fig}, yields $\alpha = 0.534$, $\beta = 1.703$, and $\sigma = 0.715$.  
Similarly for $h/W = 5/27 = 0.185$, as shown in Fig.\ \ref{fig8},  $\alpha = 0.584$, and $\beta = 1.606$, and $\sigma = 0.721$.  As shown in Fig.\ \ref{sigmaPlot}, $\sigma$ varies slowly from 0.671 at $h/W =0$ [see Eq.\ (\ref{sigmas180})] to 0.742 for large values of $h/W \ge 1$ [see Eq.\ (\ref{sigmas90})].

Integrating the Lorentz force $\phi_0 K_{H\delta}$, we find that the work done by the source of the applied field $H_z$ in moving the vortex from the corner to the position $\delta$ is 
\begin{equation}
W_H(\delta) = \phi_0 \sigma\frac{H_zW}{\Lambda} \frac{(\beta^2-\alpha^2)^{1/3}}{(\beta^2-1)^{1/3}}\Big(\frac{W}{\pi\alpha}\Big)^{1/3}\Big(\frac{3\delta}{2}\Big)^{2/3}.
\label{WHsr180}
\end{equation}

\subsubsection{$K_{H\delta}$ for the sharp  180-degree turnaround}

Starting from  Eqs.\ (\ref{SHzs180}) and (\ref{dzetadws180}), and 
doing an expansion about the origin O using Eq.\ (\ref{wexps180}), we find for $\zeta$  a short distance $\delta \ll W$ from O along the $x$ axis ($\zeta = \delta $) that the counterclockwise component of the field-induced sheet-current density is to lowest order
\begin{equation}
K_{H\delta} =\sigma \frac{H_zW}{\Lambda}\Big(\frac{W}{\pi\delta}\coth\frac{\pi}{2}\Big)^{1/2},
\label{KHdeltas180}
\end{equation}
where, with $x(u)$ obtained from Eq.\ (\ref{zetaofws180}),
\begin{equation} 
\sigma = \frac{1}{W^2}\int_{0}^\beta\frac{x(u)[x(u)-W]}{u^2}du =0.671.
\label{sigmas180}
\end{equation}

Integrating the Lorentz force $\phi_0 K_{I\delta}$, we find that the work done by the source of the current $K_I$ in moving the vortex from the corner to the position $\delta$ is 
\begin{equation}
W_H(\delta) = 2\phi_0 \sigma\frac{H_zW}{\Lambda} \Big(\frac{W\delta}{\pi}\coth\frac{\pi}{2}\Big)^{1/2}.
\label{WHs180}
\end{equation}

\subsection{Critical current for the nucleation of vortices}

We now use a London-model description of a nucleating vortex to calculate the critical current for the nucleation of vortices.  As in Ref.\ \onlinecite{Clem11}, the critical current is defined as the current that reduces the Gibbs free energy barrier at the nucleation site to zero, where the Gibbs free energy is the vortex self-energy $E_{self}(\delta)$, accounting for its interaction with all images, less the work  $W_L(\delta)$ done by the Lorentz forces $W_L(\delta)$ in moving the vortex to the position $\delta$ away from its nucleation site.  In the presence of the applied field $H_z$, this work term has contributions from both the applied current and the applied field:  $W_L(\delta) = W_I(\delta)+ W_H(\delta)$.

\subsubsection{Critical current for the sharp 90-degree turn\label{KcHzs90}}

For the sharp 90-degree turn, the Gibbs free energy becomes
\begin{equation}
G = \frac{\phi_0^2}{2\pi \mu_0 \Lambda} \ln \Big(\frac{3\delta}{\xi}\Big)-W_I(\delta)-W_H(\delta),
\label{Gs90}
\end{equation}
where the first term on the right-hand side was derived in Ref.\ \onlinecite{Clem11} and the second and third terms are from Eqs.\ (\ref{WIs90}) and (\ref{WHs90}).
Following the steps that led to Eq.\ (\ref{Kcstrip0}), we obtain  with $K_I = K_{cv}$,
\begin{eqnarray}
\delta_b^{2/3} &=& \frac{\phi_0}{2\pi \mu_0 \Lambda [K_I+\sigma(WH_z/\Lambda)]}\Big(\frac{3\pi}{4W}\Big)^{1/3},\\    
\delta_c&=&e^{3/2}\xi/3=1.49\xi,\\
K_{cv}(H_z) &= &K_{cv}(0)-\sigma\Big(\frac{WH_z}{\Lambda}\Big),
\label{KcvHzs90}
\end{eqnarray}
where $K_{cv}(0)=K_{cs}R$, $K_{cs}$ is given by Eq.\ (\ref{Kcstrip0}),
\begin{equation}
R=\frac{3}{2}\Big(\frac{\pi \xi}{4W}\Big)^{1/3},
\end{equation} 
and $\sigma = 0.742$, obtained from Eq.\ (\ref{sigmas90}).  See Fig.\ 
 \ref{Kcv&KcaFig90}(a).  For the dimensions of Fig.\ \ref{fig2}, $W = 15.5\xi$, $R = 0.555$, and $\sigma = 0.742$.

\subsubsection{Critical current for the sharp rectangular 180-degree turnaround}

For the sharp rectangular 180-degree turnaround, the Gibbs free energy becomes
\begin{equation}
G = \frac{\phi_0^2}{2\pi \mu_0 \Lambda} \ln \Big(\frac{3\delta}{\xi}\Big)-W_I(\delta)-W_H(\delta),
\label{Gsr180}
\end{equation}
where the first term on the right-hand side was derived in Ref.\ \onlinecite{Clem11} and the second and third terms are from Eqs.\ (\ref{WIsr180}) and (\ref{WHsr180}).
Following the steps as in Sec.\ \ref{KcHzs90}, we obtain  $K_{cv}(H_z)$ as in Eq.\ (\ref{KcvHzs90}), where
$K_{cv}(0)=K_{cs}R$, $K_{cs}$ is given by Eq.\ (\ref{Kcstrip0}),
\begin{equation}
R=\frac{3}{2}\frac{(\beta^2-1)^{1/3}}{(\beta^2-\alpha^2)^{1/3}}\Big(\frac{\pi \alpha \xi}{2W}\Big)^{1/3}
\end{equation} 
and $\sigma$ must be obtained numerically from  Eq.\ (\ref{sigmas90}) for a given ratio $h/W$.  For the dimensions of Fig.\ \ref{fig8}, $W = 13.5\xi$, $h = 2.5\xi$, $\alpha = 0.584$, $\beta = 1.606$,  $R = 0.545,$ and $\sigma = 0.721$.

\subsubsection{Critical current for the sharp 180-degree turnaround}

For the sharp 180-degree turnaround, the Gibbs free energy becomes
\begin{equation}
G = \frac{\phi_0^2}{2\pi \mu_0 \Lambda} \ln \Big(\frac{4\delta}{\xi}\Big)-W_I(\delta)-W_H(\delta),
\label{Gs180}
\end{equation}
where the first term on the right-hand side was derived in Ref.\ \onlinecite{Clem11} and the second and third terms are from Eqs.\ (\ref{WIs180}) and (\ref{WHs180}).
Following the steps that led to Eq.\ (\ref{Kcstrip0}), we obtain  with $K_I = K_{cv}$,
\begin{eqnarray}
\delta_b^{1/2} &=& \frac{\phi_0 [\pi\tanh(\pi/2)/W]^{1/2}}{2\pi \mu_0 \Lambda [K_I+\sigma(WH_z/\Lambda)]},\\
\delta_c&=&e^{2}\xi/4=1.85\xi,
\label{KcvHzs180}
\end{eqnarray}
and we obtain  $K_{cv}(H_z)$ as in Eq.\ (\ref{KcvHzs90}), where
$K_{cv}(0)=K_{cs}R$, $K_{cs}$ is given by Eq.\ (\ref{Kcstrip0}),
\begin{equation}
R=\Big(\frac{\pi  \xi}{W}\tanh\frac{\pi}{2}\Big)^{1/2},
\end{equation} 
and $\sigma = 0.671$, obtained from Eq.\ (\ref{sigmas180}). For the dimensions of Fig.\ \ref{fig7}, which we treat as a sharp 180-degree turnaround, $W = 15.5\xi$,  $R = 0.431$, and $\sigma =0.671$.

\end{document}